\documentclass[11pt,twoside,letterpaper]{article}
\usepackage{times,fancyhdr}
\usepackage[dvips]{graphicx}

\usepackage{amsmath}
\usepackage{amsfonts}
\usepackage[cp1251]{inputenc}

\makeatletter
\def\cleardoublepage{\clearpage\if@twoside \ifodd\c@page\else%
    \hbox{}%
    \thispagestyle{empty}%
    \newpage%
    \if@twocolumn\hbox{}\newpage\fi\fi\fi}
\makeatother

\renewcommand{\thesection}{\arabic{section}.}
\renewcommand{\thesubsection}{\thesection\arabic{subsection}.}

\begin{document}
\title{
\bfseries\scshape dynamics of
correlations \\in a system of hard spheres}
\author{\bfseries\itshape V. I. Gerasimenko\thanks{E-mail address: gerasym@imath.kiev.ua}\\
Institute of mathematics of the NAS of Ukraine\\
Kyiv, Ukraine\\ \\
\bfseries\itshape I. V. Gapyak\thanks{E-mail address: gapjak@ukr.net}\\
Taras Shevchenko National University of Kyiv,\\
Department of Mathematics and Mechanics\\
Kyiv, Ukraine}

\maketitle

\renewcommand{\headrulewidth}{0pt}

\begin{abstract}
\noindent The possible ways to describe the states of a system of many hard spheres are considered,
in particular by means of functions describing correlations of states. It is stated an approach to
the description of the evolution based on the dynamics of correlations in a system of hard spheres.
In addition, we consider another approach to describing the evolution of correlations in a system
of many hard spheres, namely, in the framework of a one-particle distribution function (correlation
function) governed by the non-Markovian Enskog kinetic equation.
\end{abstract}

\vspace{.1in}


\vspace{.08in}
                             Liouville equation, Liouville hierarchy, BBGKY hierarchy, kinetic equation.

\newpage

\tableofcontents

\newpage


\section{Introduction}
Recently, mainly in connection with the problem of the rigorous derivation of the kinetic equations
\cite{CGP97}-\cite{BPSh}, a number of articles have discussed approaches to describing the evolution
of states of a system of many hard spheres \cite{SR12}-\cite{GG21}.

As known, many-particle systems are described in terms of such notions as observables and states.
The functional for the mean value of observables determines a duality of observables and states.
As a consequence, there are two approaches to describing the evolution of a system from a finite
number of particles, namely, in terms of observables, which are governed by the Liouville equation
for  the function of observables, or in terms of states governed by the dual Liouville equation
for the probability distribution function, respectively \cite{GG21},\cite{GG18}.

An alternative approach to the description of states of a system of finitely many particles is to
describe states using functions determined by the cluster expansions of the probability distribution
function. They are cumulants (semi-invariants) of the probability distribution function and are interpreted
as correlations of the state (correlation functions). The evolution of correlation functions is governed by
the Liouville hierarchy (the von Neumann hierarchy for quantum many-particle systems \cite{GerS}-\cite{G17}).

One more approach to describing a state of many-particle systems is to describe a state by means of
a sequence of so-called reduced distribution functions (marginal distribution functions) governed by
the BBGKY (Bogolyubov--Born--Green--Kirkwood--Yvon) hierarchy \cite{CGP97}-\cite{SR12}. An alternative
approach to such a description of a state is based on sequences of functions determined by the cluster
expansions of reduced distribution functions. These functions are interpreted as the reduced correlation
functions that are governed by the hierarchy of nonlinear evolution equations (in papers \cite{BogLect},\cite{GP13}
in the case of quantum many-particle systems). The mention approaches are allowed to describe the evolution
of states of systems both with a finite and infinite number of particles.

In the paper, it is also developed an approach to the description of the evolution by means of both
reduced distribution functions and reduced correlation functions which is based on the dynamics of
correlations in a system of hard spheres. It should be emphasized that the structure of solution
expansions of the corresponding hierarchies is induced by the structure of a solution expansion of
the Liouville hierarchy for a sequence of correlation functions. We note the importance of the description
of the processes of the creation and propagation of correlations \cite{P62}, in particular, it is related
to the problem of the description of the entanglement of states in many-particle systems.

In addition in the paper, an approach to the description of the evolution of states of a hard sphere system
by means of the state of a typical particle governed by the generalized Enskog equation \cite{GG} is discussed,
or in other words, the foundations of describing the evolution of states by kinetic equations are considered.
We note that the conventional approach to the mentioned problem is based on the consideration of asymptotic
behavior (the Boltzmann--Grad asymptotic behavior \cite{G58}-\cite{PG90}) of a solution of the BBGKY hierarchy
for reduced distribution functions represented in the form of series expansions of the perturbation theory in
case of initial states specified by a one-particle distribution function without correlation functions
\cite{CGP97}-\cite{BPSh},\cite{{Bog}}.

Thus, the paper deals with the mathematical problems of the description of the evolution of many hard spheres
based on various ways of describing of the state, in particular by means of functions describing correlations
of states. Moreover, the origin of different approaches to the description of states is discussed.

\pagestyle{fancy}
\fancyhead{}
\fancyhead[EC]{V. I. Gerasimenko, I. V. Gapyak}
\fancyhead[EL,OR]{\thepage}
\fancyhead[OC]{Correlations in a hard-sphere system}
\fancyfoot{}
\renewcommand\headrulewidth{0.5pt}

\section{Dynamics of finitely many hard spheres}
The system of many hard spheres is describing in terms of observables and states.
The functional for the mean value of observables determines the duality of observables
and states, and, as a consequence, there are two equivalent ways to describe the
evolution of a system of finitely many hard spheres as the evolution of observables
governed by the Liouville equation, and as the evolution of states governed by the
dual Liouville equation (usually called the Liouville equation).
An equivalent approach adapted to describing the evolution of observables and states
of systems of both finite and infinite number of hard spheres is to describe a state
by means of a sequence of so-called reduced distribution functions (marginals) governed
by the BBGKY hierarchy of equations and of observables by means of sequences of so-called
reduced functions of observables (marginal observables) that are governed by the dual
BBGKY hierarchy of equations.

\subsection{Observables and states}
We consider a system of identical particles of a unit mass interacting as hard spheres
with a diameter of $\sigma>0$. Every particle is characterized by its phase coordinates
$(q_{i},p_{i})\equiv x_{i}\in\mathbb{R}^{3}\times\mathbb{R}^{3},\, i\geq1.$ For configurations
of such a system the following inequalities are satisfied: $|q_i-q_j|\geq\sigma,$ $i\neq j\geq1$,
i.e. the set $\mathbb{W}_n\equiv\big\{(q_1,\ldots,q_n)\in\mathbb{R}^{3n}\big||q_i-q_j|<\sigma$ for
at least one pair $(i,j):\,i\neq j\in(1,\ldots,n)\big\}$, $n>1$, is the set of forbidden configurations.

Let $C_{\gamma}$ be the space of sequences $b=(b_0,b_1,\ldots,b_n,\ldots)$ of bounded continuous functions
$b_n\in C_n$ equipped with the norm: $\|b_n\|_{C_{\gamma}}=\max_{n\geq 0}\,\frac{\gamma^{n}}{n!}\,\|b_n\|_{C_n}$,
and let $L^{1}_{n}\equiv L^{1}(\mathbb{R}^{3n}\times\mathbb{R}^{3n})$ be the space of integrable functions
that are symmetric with respect to permutations of the arguments $x_1,\ldots,x_n$, equipped with the norm: $\|f_n\|_{L^{1}(\mathbb{R}^{3n}\times\mathbb{R}^{3n})}=\int dx_1\ldots dx_n|f_n(x_1,\ldots,x_n)|$.
Hereafter the subspace of continuously differentiable functions with compact supports we will denote by
$L_{n,0}^1\subset L^1_n$ and the subspace of finite sequences of continuously differentiable functions
with compact supports let be $L_{0}^1\subset L^{1}_{\alpha}=\oplus^{\infty}_{n=0}\alpha^n L^{1}_{n}$,
where $\alpha>1$ is a real number.

For a hard-sphere system of a non-fixed, i.e. arbitrary but finite average number of identical
particles (nonequilibrium grand canonical ensemble) in the space $\mathbb{R}^{3}$, an observable
describes by the sequence $A=(A_0,A_{1}(x_1),\ldots,A_{n}(x_1,\ldots,x_n),\ldots)$ of functions
$A_{n}\in C_n$ defined on the phase spaces of the corresponding number $n$ of hard spheres.

Then the meaning of positive continuous linear functional on the space $C_{\gamma}$ is determined
the average value of an observable (the expected value or mean value of an observable). For a system
of non-fixed number of hard spheres it is defined as follows \cite{CGP97}:
\begin{eqnarray}\label{averageD}
     &&\hskip-12mm \langle A\rangle=(A,D)\doteq(I,D)^{-1}\sum\limits_{n=0}^{\infty}\frac{1}{n!}
         \,\int_{(\mathbb{R}^{3}\times\mathbb{R}^{3})^{n}}dx_{1}\ldots dx_{n}\,A_{n}\,D_{n},
\end{eqnarray}
where $D=(1,D_{1},\ldots,D_{n},\ldots)$ is a sequence of symmetric nonnegative functions
$D_{n}=D_{n}(x_1,\ldots,x_n),\,n\geq1,$ equal to zero on the set of forbidden configurations
$\mathbb{W}_n$ and the normalizing factor $(I,D)={\sum\limits}_{n=0}^{\infty}\frac{1}{n!}
\int_{(\mathbb{R}^{3}\times\mathbb{R}^{3})^{n}}dx_{1}\ldots dx_{n}D_{n}(x_1,\ldots,x_n)$ is a
grand canonical partition function.  The sequence of functions $D$  describes a state of a system
of a non-fixed number of hard spheres.

For the sequences $A\in C_{\gamma}$ and $D\in L^{1}_{\alpha}$ mean value functional \eqref{averageD}
exists and it determines a duality between observables and states.

We note that in the particular case of a system of $N<\infty$ hard spheres the observables and
states are described by the one-component sequences: $A^{(N)}=(0,\ldots,0,A_{N},0,\ldots)$ and
$D^{(N)}=(0,\ldots,0,D_{N},0,\ldots)$, respectively, and therefore, functional \eqref{averageD}
has the following representation
\begin{eqnarray*}\label{averageN}
     &&\hskip-12mm \langle A\rangle=(A,D)\doteq\frac{1}{N!}(I,D)^{-1}
         \,\int_{(\mathbb{R}^{3}\times\mathbb{R}^{3})^{N}}dx_{1}\ldots dx_{N}\,A_{N}\,D_{N},
\end{eqnarray*}
where $(I,D)=\frac{1}{N!}\int_{(\mathbb{R}^{3}\times\mathbb{R}^{3})^{N}}dx_{1}\ldots dx_{N}D_{N}$
is the normalizing factor (canonical partition function), and it is usually assumed that the
normalization condition $\int_{(\mathbb{R}^{3}\times\mathbb{R}^{3})^{N}}dx_{1}\ldots dx_{N}D_{N}=1$
holds.

The function $D_{N}(x_1,\ldots,x_N)$, which describes all possible states of a system of $N$ hard spheres,
is called a probability distribution function, since the expression $(I,D)^{-1}D_{N}(x_1,\ldots,x_N)dx_{1}\ldots dx_{N}$
is the probability of finding the phase states of the $1st,\ldots,Nth$ hard sphere in the phase volumes
$dx_{1},\ldots, dx_{N}$ centered at the phase points $x_1,\ldots,x_N$, respectively.

If at initial instant an observable specified by the sequence
$A(0)=(A_0,A_{1}^{0}(x_1),\ldots,$ $A_{n}^{0}(x_1,\ldots,x_n),\ldots)$, then the evolution
of observables $A_{n}(t),\,n\geq1$, i.e. the sequence $A(t)=(A_0,A_{1}(t,x_1),\ldots,A_{n}(t,x_1,\ldots,x_n),\ldots)$
is determined by the following the one-parameter mapping $S(t)=\oplus_{n=0}^\infty S_n(t)$:
\begin{eqnarray}\label{sH}
    &&A(t)=S(t)A(0),
\end{eqnarray}
which is defined on every the space $C_n\equiv C(\mathbb{R}^{3n}\times(\mathbb{R}^{3n}\setminus \mathbb{W}_n))$
by means of the phase trajectories of a hard-sphere system, which are defined almost everywhere on
the phase space $\mathbb{R}^{3n}\times(\mathbb{R}^{3n}\setminus \mathbb{W}_n)$, namely, beyond of the set
$\mathbb{M}_{n}^0$ of the zero Lebesgue measure, as follows
\begin{eqnarray} \label{Sspher}
  &&\hskip-5mm(S_{n}(t)b_{n})(x_{1},\ldots,x_{n})\equiv S_{n}(t,1,\ldots,n)b_{n}(x_{1},\ldots,x_{n})\doteq\\
  &&\begin{cases}
         b_{n}(X_{1}(t,x_{1},\ldots,x_{n}),\ldots,X_{n}(t,x_{1},\ldots,x_{n})),\\
         \hskip+45mm\mathrm{if}\,(x_{1},\ldots,x_{n})\in(\mathbb{R}^{3n}\times(\mathbb{R}^{3n}\setminus\mathbb{W}_{n})),\\
         0, \hskip+42mm \mathrm{if}\,(q_{1},\ldots,q_{n})\in\mathbb{W}_{n},
                    \end{cases}\nonumber
\end{eqnarray}
where for $t\in\mathbb{R}$ the function $X_{i}(t)$ is a phase trajectory of $ith$ particle
constructed in \cite{CGP97} and the set $\mathbb{M}_{n}^0$ consists of the phase space points which
are specified such initial data that during the evolution generate multiple collisions, i.e. collisions
of more than two particles, more than one two-particle collision at the same instant and infinite number
of collisions within a finite time interval \cite{CGP97},\cite{PG90}.

On the space $C_n$ one-parameter mapping (\ref{Sspher}) is an isometric $\ast$-weak continuous
group of operators, i.e. it is a $C_{0}^{\ast}$-group. For the group of evolution operators (\ref{Sspher})
the Duhamel equation holds
\begin{eqnarray*}\label{DuamN}
    &&\hskip-5mm S_n(t,1,\ldots,n)b_n=\\
    &&\prod\limits_{i=1}^nS_1(t,i)b_n+\int\limits_0^td\tau \prod\limits_{i=1}^{n}S_{1}(t-\tau,i)
       \sum\limits_{j_1<j_2=1}^{n}\mathcal{L}_{\mathrm{int}}(j_1,j_2)S_{n}(\tau,1,\ldots,n)b_{n}=\nonumber\\
    &&\prod\limits_{i=1}^nS_1(t,i)b_n+\int\limits_0^td\tau S_{n} (t-\tau,1,\ldots,n)
       \sum\limits_{j_{1}<j_{2}=1}^{n}\mathcal{L}_{\mathrm{int}}(j_{1},j_{2})
       \prod\limits_{i=1}^{n}S_{1}(\tau,i)b_{n},\nonumber
\end{eqnarray*}
where for $t>0$ the operator $\mathcal{L}_{\mathrm{int}}(j_{1},j_{2})$ is defined by the formula
\begin{eqnarray}\label{Lint}
     &&\mathcal{L}_{\mathrm{int}}(j_{1},j_{2})b_{n}\doteq
        \sigma^2\int_{\mathbb{S}_{+}^2}d\eta\langle\eta,(p_{j_{1}}-p_{j_{2}})\rangle
        \big(b_n(x_1,\ldots,p_{j_{1}}^\ast,q_{j_{1}},\ldots,\\
     &&\hskip+21mm p_{j_{2}}^\ast,q_{j_{2}},\ldots,x_n)-b_n(x_1,\ldots,x_n)\big)\delta(q_{j_{1}}-q_{j_{2}}+\sigma\eta).\nonumber
\end{eqnarray}
In definition (\ref{Lint}) the symbol $\langle \cdot,\cdot \rangle$ means a scalar product, the symbol $\delta$ denotes
the Dirac measure, $\mathbb{S}_{+}^{2}\doteq\{\eta\in\mathbb{R}^{3}\big|\left|\eta\right|=1\langle\eta,(p_1-p_{2})\rangle>0\}$
and the momenta $p_{i}^\ast,p_{j}^\ast$ are determined by the equalities:
\begin{eqnarray*}\label{momenta}
     &&p_i^\ast\doteq p_i-\eta\left\langle\eta,\left(p_i-p_{j}\right)\right\rangle, \\
     &&p_{j}^\ast\doteq p_{j}+\eta\left\langle\eta,\left(p_i-p_{j}\right)\right\rangle \nonumber.
\end{eqnarray*}
If $t<0$, the operator $\mathcal{L}_{\mathrm{int}}(j_{1},j_{2})$ is defined by the corresponding
expression \cite{CGP97}.

Thus, the infinitesimal generator $\mathcal{L}_{n}$ of the group of operators (\ref{Sspher}) has the structure
\begin{eqnarray}\label{L}
     &&\mathcal{L}_{n}b_{n}\doteq\sum\limits_{j=1}^{n}\mathcal{L}(j)b_{n}+
        \sum\limits_{j_{1}<j_{2}=1}^{n}\mathcal{L}_{\mathrm{int}}(j_{1},j_{2})b_{n},
\end{eqnarray}
where the Liouville operator of free motion
$\mathcal{L}(j)\doteq\langle p_j,\frac{\partial}{\partial q_j}\rangle$ defined on
the set $C_{n,0}$, we had denoted by the symbol $\mathcal{L}(j)$.

If $A(0)\in C_{\gamma}$, the sequence (\ref{sH}) is a unique solution of the Cauchy problem for the sequence
of the weak formulation of the Liouville equations
\begin{eqnarray}
  \label{H-N1}
     &&\frac{\partial}{\partial t}A(t)=\mathcal{L}A(t),\\
  \label{H-N12}
     &&A(t)|_{t=0}=A(0),
\end{eqnarray}
where the operator $\mathcal{L}=\oplus_{n=0}^\infty\mathcal{L}_n$ is defined by formula (\ref{L}).

Taking into account the equality $(I,D(0))=(I,S^\ast(t)D(0))$, and of the validity for functional
\eqref{averageD} the following representations:
\begin{eqnarray}\label{os}
     &&\hskip-12mm (A(t),D(0))=(I,D(0))^{-1}\sum\limits_{n=0}^{\infty}\frac{1}{n!}
         \,\int_{(\mathbb{R}^{3}\times\mathbb{R}^{3})^{n}}dx_{1}\ldots dx_{n}\,
         S_{n}(t)A_{n}^0\,D_{n}^0=\\
     &&(I,S^\ast(t)D(0))^{-1}\sum\limits_{n=0}^{\infty}\frac{1}{n!}\,
         \int_{(\mathbb{R}^{3}\times\mathbb{R}^{3})^{n}}dx_{1}\ldots dx_{n}\,
         A_{n}^0\,S^\ast_n(t)D_{n}^0\equiv \nonumber\\
     &&(I,D(t))^{-1}(A(0),D(t)),\nonumber
\end{eqnarray}
where the adjoint group of operators $S_n^\ast(t)$ to group of operators (\ref{Sspher}) is defined
on the space of integrable functions $L^1_n$
\begin{eqnarray}\label{ad}
   &&S_n^\ast(t)=S_n(-t),
\end{eqnarray}
then, as a result, we can describe the evolution of many hard spheres within the evolution of states.

On the space $L_n^1$ the one-parameter mapping defined by formula (\ref{ad}) is an isometric strong continuous
group of operators. Indeed, $\big\|S_n^\ast(t)\big\|=1$.

We note that the group of operators (\ref{ad}) satisfies the Duhamel equation
\begin{eqnarray*}\label{DuamN_1}
    &&\hskip-8mm S_n^\ast(t,1,\ldots,n)=\\
    &&\prod\limits_{i=1}^{n}S_1^\ast(t,i)+\int\limits_0^t d\tau \prod\limits_{i=1}^{n}S_{1}^\ast(t-\tau,i)
       \sum\limits_{j_{1}<j_{2}=1}^{n}\mathcal{L}_{\mathrm{int}}^\ast(j_{1},j_{2})S_{n}^\ast(\tau,1,\ldots,n))=\\
    &&\prod\limits_{i=1}^{n}S_1^\ast(t,i)+\int\limits_0^td\tau S_{n}^\ast(t-\tau,1,\ldots,n)
       \sum\limits_{j_{1}<j_{2}=1}^{n}\mathcal{L}_{\mathrm{int}}^\ast(j_{1},j_{2})
       \prod\limits_{i=1}^{n}S_{1}^\ast(\tau,i),\nonumber
\end{eqnarray*}
where for $t>0$ the operator $\mathcal{L}_{\mathrm{int}}^\ast(j_{1},j_{2})$ is defined by the formula
\begin{eqnarray}\label{bLint}
     &&\hskip-9mm \mathcal{L}_{\mathrm{int}}^\ast(j_{1},j_{2})f_{n}
        \doteq\sigma^2\int_{\mathbb{S}_{+}^2}d\eta\langle\eta,(p_{j_{1}}-p_{j_{2}})\rangle
        f_n(x_1,\ldots,p_{j_{1}}^*,q_{j_{1}},\ldots,\\
     &&p_{j_{2}}^*,q_{j_{2}},\ldots,x_n)\delta(q_{j_{1}}-q_{j_{2}}+\sigma\eta)-
        f_n(x_1,\ldots,x_n)\delta(q_{j_{1}}-q_{j_{2}}-\sigma\eta)\big).\nonumber
\end{eqnarray}
In formula (\ref{bLint}) the notations similar to (\ref{Lint}) are used.

Hence the infinitesimal generator $\mathcal{L}_{n}^\ast$ of the group of operators $S_n^\ast(t)$
has the structure
\begin{eqnarray}\label{Lstar}
    &&\mathcal{L}_{n}^\ast f_{n}\doteq\sum\limits_{j=1}^{n}\mathcal{L}^\ast(j)f_{n}+
        \sum\limits_{j_{1}<j_{2}=1}^{n}\mathcal{L}_{\mathrm{int}}^\ast(j_{1},j_{2})f_{n},
\end{eqnarray}
where the Liouville operator of free motion
$\mathcal{L}^{\ast}(j)\doteq-\langle p_j,\frac{\partial}{\partial q_j}\rangle$ defined on
the subspace $L_{n,0}^1\subset L^1_n$, we had denoted by the symbol $\mathcal{L}^{\ast}(j)$.

In view of the validity of equality (\ref{os}) the evolution of all possible states, i.e. the
sequence $D(t)=(1,D_{1}(t),\ldots,$ $D_{n}(t),\ldots)\in L^{1}_{\alpha}$ of the probability
distribution functions $D_{n}(t),\,n\geq1$, is determined by the formula
\begin{eqnarray}\label{rozv_fon-N}
    &&D(t)=S^\ast(t)D(0),
\end{eqnarray}
where the one-parameter family of operators $S^\ast(t)=\oplus_{n=0}^{\infty}S^\ast_{n}(t)$,
is defined as above.

If $D(0)\in L^{1}_{\alpha}$, the sequence of distribution functions defined by formula
(\ref{rozv_fon-N}) is a unique solution of the Cauchy problem for the sequence of the weak
formulation of the dual Liouville equation for states (known as the Liouville equation)
\begin{eqnarray}
  \label{vonNeumannEqn}
     &&\frac{\partial}{\partial t}D(t)=\mathcal{L}^\ast D(t),\\
  \label{F-N12}
     &&D(t)|_{t=0}=D(0),
\end{eqnarray}
where the generator $\mathcal{L}^\ast=\oplus^{\infty}_{n=0}\mathcal{L}^\ast_{n}$ of the dual
Liouville equations \eqref{vonNeumannEqn} is the adjoint operator to generator \eqref{L} of
the Liouville equation \eqref{H-N1} in the sense of functional \eqref{averageD}, i.e. it is
defined by formula \eqref{Lstar}.


\subsection{Reduced functions of observables and states}
For the description of a system of hard spheres of both finite and infinite number
of particles another approach to describing observables and states is used, which
is equivalent to the approach formulated above in the case of systems of finitely
many hard spheres \cite{CGP97},\cite{Bog}.

Indeed, for a system of finitely many particles mean value functional \eqref{averageD}
can be represented in one more form
\begin{eqnarray}\label{avmar}
       &&\hskip-12mm\langle A\rangle=(I,D)^{-1}\sum\limits_{n=0}^{\infty}\frac{1}{n!}\,
         \int_{(\mathbb{R}^{3}\times\mathbb{R}^{3})^{n}}dx_{1}\ldots dx_{n}\,A_{n}\,D_{n}=\\
       &&\sum\limits_{s=0}^{\infty}\frac{1}{s!}\,
         \int_{(\mathbb{R}^{3}\times\mathbb{R}^{3})^{s}}dx_{1}\ldots dx_{s}\,B_{s}(x_1,\ldots,x_s)\,
         F_{s}(x_1,\ldots,x_s),\nonumber
\end{eqnarray}
where, for the description of observables and states, the sequence of so-called reduced functions of
observables $B=(B_0,B_{1}(x_1),\ldots,B_{s}(x_1,\ldots,x_s),\ldots)$ (other used terms: marginal or
$s$-particle observable \cite{GG21}) was introduced and the sequence of so-called reduced distribution
functions $F=(1,F_{1}(x_1),\ldots,F_{s}(x_1,\ldots,x_s),\ldots)$ (other used terms: marginals \cite{SR12},
\cite{BPSh}, truncated or $s$-particle distribution function \cite{Bog}), respectively. Thus, the reduced
functions of observables are defined by means functions of observables by the following expansions \cite{GG18}:
\begin{eqnarray}\label{mo}
    &&\hskip-12mm B_{s}(x_1,\ldots,x_s)\doteq\sum_{n=0}^s\,\frac{(-1)^n}{n!}\sum_{j_1\neq\ldots\neq j_{n}=1}^s
          A_{s-n}((1,\ldots,s)\setminus(j_1,\ldots,j_{n})), \quad s\geq 1,
\end{eqnarray}
and the reduced distribution functions are defined by means of probability distribution functions as follows \cite{CGP97}
\begin{eqnarray}\label{ms}
     &&\hskip-8mm F_{s}(x_1,\ldots,x_s)\doteq\\
     &&\hskip+2mm (I,D)^{-1}\sum\limits_{n=0}^{\infty}\frac{1}{n!}
         \,\int_{(\mathbb{R}^{3}\times\mathbb{R}^{3})^{n}}dx_{s+1}\ldots dx_{s+n}\,D_{s+n}(x_1,\ldots,x_{s+n}),
         \quad s\geq 1.\nonumber
\end{eqnarray}

We emphasize that the possibility of describing states within the framework of reduced distribution
functions naturally arises as a result of dividing the series in expression \eqref{averageD}
by the series of the normalization factor, i.e. in consequence of redefining of mean value
functional \eqref{avmar}.

If initial state specified by the sequence of reduced distribution functions
$F(0)=(1,F_{1}^{0}(x_1),\ldots,F_{n}^{0}(x_1,\ldots,x_n),\ldots)\in L^{1}_{\alpha}$, then the evolution
of all possible states, i.e. the sequence $F(t)=(1,F_{1}(t,x_1),\ldots,F_{s}(t,x_1,\ldots,x_s),\ldots)$
of the reduced distribution functions $F_{s}(t),\,s\geq1$, is determined by the following series
expansions \cite{GerRS}:
\begin{eqnarray}\label{RozvBBGKY}
   &&\hskip-8mm F_{s}(t,x_1,\ldots,x_s)=\sum\limits_{n=0}^{\infty}\frac{1}{n!}\,
       \int_{(\mathbb{R}^{3}\times\mathbb{R}^{3})^{n}}dx_{s+1}\ldots dx_{s+n}\,
       \mathfrak{A}_{1+n}(t,\{1,\ldots,s\},\\
   &&\hskip+8mm  s+1,\ldots,{s+n})F_{s+n}^{0}(x_1,\ldots,x_{s+n}),\quad s\geq1,\nonumber
\end{eqnarray}
where the generating operator
\begin{eqnarray}\label{cumulant1+n}
   &&\hskip-8mm \mathfrak{A}_{1+n}(t,\{1,\ldots,s\},s+1,\ldots,{s+n})=\\
   &&\hskip+8mm \sum\limits_{\mathrm{P}\,:(\{1,\ldots,s\},s+1,\ldots,{s+n})=
      {\bigcup\limits}_i X_i}(-1)^{|\mathrm{P}|-1}(|\mathrm{P}|-1)!
      \prod_{X_i\subset\mathrm{P}}S^\ast_{|\theta(X_i)|}(t,\theta(X_i))\nonumber
\end{eqnarray}
is the $(1+n)th$-order cumulant of the groups of operators \eqref{rozv_fon-N} \cite{GerRS}.
In expansion \eqref{cumulant1+n} the symbol $\{1,\ldots,s\}$ is a set consisting of one element
$(1,\ldots,s)$, i.e. $|\{1,\ldots,s\}|=1$, ${\sum\limits}_\mathrm{P}$ means the sum over
all possible partitions $\mathrm{P}$ of the set $(\{1,\ldots,s\},s+1,\ldots,{s+n})$ into
$|\mathrm{P}|$ nonempty mutually disjoint subsets $X_i\subset(\{1,\ldots,s\},s+1,\ldots,{s+n})$
and $\theta$ is the declusterization mapping: $\theta(\{1,\ldots,s\},s+1,\ldots,{s+n})\doteq(1,\ldots,s+n)$.
The simplest examples of cumulants \eqref{cumulant1+n} of groups of operators \eqref{rozv_fon-N}
have the form
\begin{eqnarray*}
   &&\mathfrak{A}_{1}(t,\{1,\ldots,s\})=S^\ast_{s}(t,1,\ldots,s),\\
   &&\mathfrak{A}_{1+1}(t,\{1,\ldots,s\},s+1)=S^\ast_{s+1}(t,1,\ldots,s+1)-S^\ast_{s}(t,1,\ldots,s)S^\ast_{1}(t,s+1).
\end{eqnarray*}
If $F(0)\in L^{1}_{\alpha}=\oplus_{n=0}^{\infty}\alpha^{n}L^{1}_{n}$, series \eqref{RozvBBGKY}
converges on the norm of the space $L^{1}_{\alpha}$ provided that $\alpha>e$. The parameter $\alpha$
can be interpreted as the magnitude inverse to the average number of hard spheres.

We note that the method of constructing the reduced distribution functions \eqref{RozvBBGKY}
is based on the application of cluster expansions to the generating operators \eqref{rozv_fon-N}
of series \eqref{ms}, as a result of which the generating operators of series \eqref{RozvBBGKY}
are the corresponding-order cumulants of the groups of operators $S^\ast(t)$ \cite{CGP97},\cite{GerRS}.

If $F(0)\in\oplus_{n=0}^{\infty}\alpha^{n}L^{1}_{n}$ and $\alpha>e$, then for $t\in\mathbb{R}$
the sequence of reduced distribution functions \eqref{RozvBBGKY} is a unique solution of the Cauchy
problem of the BBGKY hierarchy \cite{CGP97},\cite{Bog}:
\begin{eqnarray}
 \label{BBGKY}
   &&\hskip-8mm\frac{\partial}{\partial t}F_{s}(t,x_1,\ldots,x_s)=\mathcal{L}^{\ast}_{s}F_{s}(t,x_1,\ldots,x_s)+ \\
   &&\hskip+8mm\sum\limits_{j=1}^{s}\int_{(\mathbb{R}^{3}\times\mathbb{R}^{3})}dx_{s+1}
     \mathcal{L}^{\ast}_{\mathrm{int}}(j,s+1)F_{s+1}(t,x_1,\ldots,x_{s+1}),\nonumber\\  \nonumber \\
 \label{BBGKYi}
   &&\hskip-8mm F_{s}(t,x_1,\ldots,x_s)\mid_{t=0}=F_{s}^{0}(x_1,\ldots,x_s),\quad s\geq 1,
\end{eqnarray}
where we used notations accepted in formula \eqref{Lstar}, i.e. for $t\geq0$ the Liouville operator $\mathcal{L}^{\ast}_{s}$
is defined in \cite{CGP97} and the equality holds
\begin{eqnarray*}
   &&\hskip-8mm\sum_{j=1}^{s}\int_{\mathbb{R}^{3}\times\mathbb{R}^{3}}dx_{s+1}
     \mathcal{L}^{*}_{\mathrm{int}}(j,s+1)F_{s+1}(t)\doteq\\
   &&\sigma^{2}\sum\limits_{i=1}^s\int_{\mathbb{R}^3\times\mathbb{S}_{+}^{2}}d p_{s+1}d\eta\,\langle\eta,(p_i-p_{s+1})\rangle
      \big(F_{s+1}(t,x_1,\ldots,q_i,p_i^{*},\ldots,\\
   &&x_s,q_i-\sigma\eta,p_{s+1}^{*})-F_{s+1}(t,x_1,\ldots,x_s,q_i+\sigma\eta,p_{s+1})\big),
\end{eqnarray*}
and for $t\leq0$, the generator of the BBGKY hierarchy \eqref{BBGKY} is defined by the corresponding expression \cite{CGP97}.
Sequences of functions from the space $L^1_{\alpha}$ describe the state of a finitely many-particle system, because in this
case the average number of hard spheres $\langle N\rangle=\,\int_{(\mathbb{R}^{3}\times\mathbb{R}^{3})}dx_{1}\,F_{1}(t,x_1)$
is finite.

We note that traditionally \cite{CGP97},\cite{BPSh},\cite{Bog} the reduced distribution functions
are represented by means of the perturbation theory series of the BBGKY hierarchy \eqref{BBGKY}
\begin{eqnarray*}\label{iter}
   &&\hskip-5mm F_s(t,x_1,\ldots,x_s)=\\
   &&\hskip-2mm \sum\limits_{n=0}^{\infty}\,\int\limits_{0}^{t}dt_{1}\ldots
       \int\limits_{0}^{t_{n-1}}dt_{n}\int_{(\mathbb{R}^{3}\times\mathbb{R}^{3})^{n}}dx_{s+1}\ldots dx_{s+n}\,S^\ast_s(t-t_{1})
       \sum\limits_{j_1=1}^{s}\mathcal{L}^{\ast}_{\mathrm{int}}(j_1,s+1))\times
       \nonumber\\
   &&\hskip-2mm S^\ast_{s+1}(t_1-t_2)\ldots S^\ast_{s+n-1}(t_{n-1}-t_n)
       \sum\limits_{j_n=1}^{s+n-1}\mathcal{L}^{\ast}_{\mathrm{int}}(j_n,s+n))
       S^\ast_{s+n}(t_{n})F_{s+n}^0(x_1,\ldots,       \nonumber\\
   &&\hskip-2mm x_{s+n}),\quad s\geq1,\nonumber
\end{eqnarray*}
where we used notations accepted in formula \eqref{bLint}. The nonperturbative series expansion
for reduced distribution functions \eqref{RozvBBGKY} is represented in the form of the perturbation
theory series for suitable interaction potentials and initial data as a result of the employment
of analogs of the Duhamel equation to cumulants \eqref{cumulant1+n} of the groups of operators \eqref{rozv_fon-N}.

We remark that, if initial observable (\ref{mo}) specified by the sequence of reduced observables
$B(0)=(B_0,B_1^{0}(x_1),\ldots,B_s^{0}(x_1,\ldots,x_s),\ldots)\in\mathcal{C}_{\gamma}$, then the evolution
of observables, i.e. the sequence $B(t)=(B_0,B_1(t,x_1),\ldots,B_s(t,x_1,\ldots,x_s),\ldots)$
of the reduced observables $B_{s}(t),\,s\geq1$, is determined by the following series
expansions \cite{GG18}:
\begin{eqnarray}\label{sdh}
   &&\hskip-12mm B_{s}(t,x_1,\ldots,x_s)=\sum_{n=0}^s\,\frac{1}{n!}\sum_{j_1\neq\ldots\neq j_{n}=1}^s
      \mathfrak{A}_{1+n}\big(t,\{(1,\ldots,s)\setminus (j_1,\ldots,j_{n})\},\\
   &&\hskip-7mm (j_1,\ldots,j_{n})\big)\,
      B_{s-n}^{0}(x_1,\ldots,x_{j_1-1},x_{j_1+1},\ldots,x_{j_n-1},x_{j_n+1},\ldots,x_s),\quad s\geq1.\nonumber
\end{eqnarray}
The generating operators of expansions (\ref{sdh}) is the $(1+n)th$-order cumulant
of groups of operators (\ref{Sspher}) defined by the following expansion:
\begin{eqnarray*}\label{cumulant}
    &&\hskip-7mm\mathfrak{A}_{1+n}(t,\{(1,\ldots,s)\setminus (j_1,\ldots,j_{n})\},(j_1,\ldots,j_{n}))\doteq\\
    &&\hskip-7mm\sum\limits_{\mathrm{P}:\,(\{(1,\ldots,s)\setminus (j_1,\ldots,j_{n})\},(j_1,\ldots,j_{n}))={\bigcup}_i X_i}
       (-1)^{\mathrm{|P|}-1}({\mathrm{|P|}-1})!\prod_{X_i\subset \mathrm{P}}
       S_{|\theta(X_i)|}(t,\theta(X_i)),\quad n\geq0,
\end{eqnarray*}
where the symbol ${\sum}_\mathrm{P}$ means the sum over all possible partitions $\mathrm{P}$ of the
set $(1,\ldots,n)$ into $|\mathrm{P}|$ nonempty mutually disjoint subsets $X_i\subset(1,\ldots,n)$.
This sequence is a unique solution of the Cauchy problem of the weak formulation of the dual BBGKY
hierarchy for hard spheres \cite{GShZ},\cite{GG18}:
\begin{eqnarray}\label{dh}
   &&\hskip-9mm\frac{\partial}{\partial t}B_{s}(t,x_1,\ldots,x_s)=\big(\sum\limits_{j=1}^{s}\mathcal{L}(j)+
      \sum\limits_{j_1<j_{2}=1}^{s}\mathcal{L}_{\mathrm{int}}(j_1,j_{2})\big)B_{s}(t,x_1,\ldots,x_s)+\\
   &&\hskip-5mm+\sum_{j_1\neq j_{2}=1}^s
      \mathcal{L}_{\mathrm{int}}(j_1,j_{2})B_{s-1}(t,x_1,\ldots,x_{j_1-1},x_{j_1+1},\ldots,x_s),\nonumber\\
      \nonumber\\
\label{dhi}
   &&\hskip-9mm B_{s}(t,x_1,\ldots,x_s)_{\mid t=0}=B_{s}^{0}(x_1,\ldots,x_s),\quad s\geq1,
\end{eqnarray}
where it is used notations accepted in formula (\ref{L}).

Thus, there exist two approaches to the description of the evolution of many hard spheres, namely,
within the framework of observables that are governed by the dual BBGKY hierarchy (\ref{dh}) for
reduced functions of observables, or in terms of states governed by the BBGKY hierarchy (\ref{BBGKY})
for the reduced distribution functions, respectively. For a system of finitely many hard spheres, these
hierarchies are equivalent to the Liouville equation for observables and to the Liouville equation for
states (the dual Liouville equation), respectively.


\section{Dynamics of correlations of a hard-sphere system}
An alternative approach to the description of states of a hard-sphere system of finitely
many particles is given by means of functions determined by the cluster expansions of the
probability distribution functions. They are interpreted as correlation functions (cumulants
of probability distribution functions).

\subsection{Correlation functions}
We introduce the sequence of correlation functions $g(t)=(1,g_{1}(t,x_1),\ldots,g_{s}(t,x_1,$ $\ldots,x_s),\ldots)$
by means of the cluster expansions of the probability distribution functions
$D(t)=(1,D_{1}(t,x_1),\ldots,D_{n}(t,x_1,\ldots,x_n),\ldots)$, defined on the set of allowed
configurations $\mathbb{R}^{3n}\setminus \mathbb{W}_n$ as follows:
\begin{eqnarray}\label{D_(g)N}
    &&\hskip-12mm D_{n}(t,x_1,\ldots,x_n)= g_{n}(t,x_1,\ldots,x_n)+\sum\limits_{\mbox{\scriptsize $\begin{array}{c}\mathrm{P}:
    (x_1,\ldots,x_n)=\bigcup_{i}X_{i},\\|\mathrm{P}|>1 \end{array}$}}
        \prod_{X_i\subset \mathrm{P}}g_{|X_i|}(t,X_i),\\
    &&\hskip-12mm n\geq1,\nonumber
\end{eqnarray}
where ${\sum\limits}_{\mathrm{P}:(x_1,\ldots,x_n)=\bigcup_{i} X_{i},\,|\mathrm{P}|>1}$ is the
sum over all possible partitions $\mathrm{P}$ of the set of the arguments $(x_1,\ldots,x_n)$
into $|\mathrm{P}|>1$ nonempty mutually disjoint subsets $X_i\subset(x_1,\ldots,x_n)$.

On the set $\mathbb{R}^{3n}\setminus \mathbb{W}_n$ solutions of recursion relations \eqref{D_(g)N}
are given by the following expansions:
\begin{eqnarray}\label{gfromDFB}
   &&\hskip-8mm g_{s}(t,x_1,\ldots,x_s)=D_{s}(t,x_1,\ldots,x_s)+\nonumber\\
   &&\sum\limits_{\mbox{\scriptsize $\begin{array}{c}\mathrm{P}:(x_1,\ldots,x_s)=
       \bigcup_{i}X_{i},\\|\mathrm{P}|>1\end{array}$}}(-1)^{|\mathrm{P}|-1}(|\mathrm{P}|-1)!\,
       \prod_{X_i\subset \mathrm{P}}D_{|X_i|}(t,X_i), \quad s\geq1.
\end{eqnarray}
The structure of expansions \eqref{gfromDFB} is such that the correlation functions can be treated
as cumulants (semi-invariants) of the probability distribution functions \eqref{rozv_fon-N}.

Thus, correlation functions \eqref{gfromDFB} are to enable to describe of the evolution of states
of finitely many hard spheres by the equivalent method in comparison with the probability distribution
function, namely within the framework of dynamics of correlations \cite{GerS},\cite{GP11}.

If initial state described by the sequence $g(0)=(1,g_{1}^{0}(1),\ldots,g_{n}^{0}(x_1,\ldots,x_n),\ldots)$,
of correlation functions $g_{n}^{0}\in L^{1}_{n},\,n\geq1,$ then the evolution of all possible states,
i.e. the sequence $g(t)=(1,g_{1}(t,x_1),\ldots,g_{s}(t,x_1,\ldots,x_s),\ldots)$ of the correlation
functions $g_{s}(t),\,s\geq1$, is determined by the following group of nonlinear operators \cite{GP11}:
\begin{eqnarray}\label{ghs}
   &&\hskip-8mm g_{s}(t,x_1,\ldots,x_s)=\mathcal{G}(t;1,\ldots,s\mid g(0))\doteq\\
   &&\sum\limits_{\mathrm{P}:\,(1,\ldots,s)=\bigcup_j X_j}
      \mathfrak{A}_{|\mathrm{P}|}(t,\{X_1\},\ldots,\{X_{|\mathrm{P}|}\})
      \prod_{X_j\subset \mathrm{P}}g_{|X_j|}^{0}(X_j),\quad s\geq1,\nonumber
\end{eqnarray}
where $\sum_{\mathrm{P}:\,(1,\ldots,s)=\bigcup_j X_j}$ is the sum over all possible partitions
$\mathrm{P}$ of the set $(1,\ldots,s)$ into $|\mathrm{P}|$ nonempty mutually disjoint subsets
$X_j$, the set $(\{X_1\},\ldots,\{X_{|\mathrm{P}|}\})$ consists from elements of which are
subsets $X_j\subset (1,\ldots,s)$, i.e. $|(\{X_1\},\ldots,\{X_{|\mathrm{P}|}\})|=|\mathrm{P}|$.
The generating operator $\mathfrak{A}_{|\mathrm{P}|}(t)$ in expansion \eqref{ghs} is the
$|\mathrm{P}|th$-order cumulant of the groups of operators \eqref{rozv_fon-N} which is defined
by the expansion
\begin{eqnarray}\label{cumulantP}
   &&\hskip-8mm \mathfrak{A}_{|\mathrm{P}|}(t,\{X_1\},\ldots,\{X_{|\mathrm{P}|}\})\doteq\\
   && \sum\limits_{\mathrm{P}^{'}:\,(\{X_1\},\ldots,\{X_{|\mathrm{P}|}\})=
      \bigcup_k Z_k}(-1)^{|\mathrm{P}^{'}|-1}({|\mathrm{P}^{'}|-1})!
      \prod\limits_{Z_k\subset\mathrm{P}^{'}}S^\ast_{|\theta(Z_{k})|}(t,\theta(Z_{k})),\nonumber
\end{eqnarray}
where $\theta$ is the declusterization mapping: $\theta(\{X_1\},\ldots,\{X_{|\mathrm{P}|}\})\doteq(1,\ldots,s)$.
The simplest examples of correlation operators \eqref{ghs} are given by the following expansions:
\begin{eqnarray*}
   &&g_{1}(t,x_1)=\mathfrak{A}_{1}(t,1)g_{1}^{0}(x_1),\\
   &&g_{2}(t,x_1,x_2)=\mathfrak{A}_{1}(t,\{1,2\})g_{2}^{0}(x_1,x_2)+
     \mathfrak{A}_{2}(t,1,2)g_{1}^{0}(x_1)g_{1}^{0}(x_2).
\end{eqnarray*}

Thus, the cumulant nature of correlation functions induces the cumulant structure
of a one-parametric mapping \eqref{ghs}.

In particular, in the absence of correlations between hard spheres at the initial moment,
known as the initial states satisfying the chaos condition \cite{CGP97}-\cite{Sp91}, the sequence
of the initial correlation functions has the form $g^{(c)}(0)=(0,g_{1}^{0}(x_1),0,\ldots,0,\ldots)$
(in terms of a sequence of the probability distribution functions it means that
$D^{(c)}(0)=(1,D_{1}^0(x_1),$ $D_{1}^0(x_1)D_{1}^0(x_2)\mathcal{X}_{\mathbb{R}^{6}\setminus \mathbb{W}_2},\ldots,
\prod^n_{i=1} D_{1}^0(x_i)\mathcal{X}_{\mathbb{R}^{3n}\setminus \mathbb{W}_n},\ldots)$,
where $\mathcal{X}_{\mathbb{R}^{3n}\setminus \mathbb{W}_{n}}$ is the Heaviside step function of
allowed configurations of $n$ hard spheres).
In this case for $(x_1,\ldots,x_s)\in\mathbb{R}^{3s}\times(\mathbb{R}^{3s}\setminus \mathbb{W}_s)$
expansions \eqref{ghs} are represented as follows:
\begin{eqnarray}\label{gth}
   &&g_{s}(t,x_1,\ldots,x_s)=\mathfrak{A}_{s}(t,1,\ldots,s)\,\prod\limits_{i=1}^{s}g_{1}^{0}(x_i),\quad s\geq1,
\end{eqnarray}
where $\mathfrak{A}_{s}(t)$ is the $sth$-order cumulant of groups of operators \eqref{rozv_fon-N} defined
by the expansion
\begin{eqnarray}\label{cumcp}
   &&\hskip-8mm\mathfrak{A}_{s}(t,1,\ldots,s)=\sum\limits_{\mathrm{P}:\,(1,\ldots,s)=
       \bigcup_i X_i}(-1)^{|\mathrm{P}|-1}({|\mathrm{P}|-1})!
      \prod\limits_{X_i\subset\mathrm{P}}S^\ast_{|X_i|}(t,X_i),
\end{eqnarray}
and it was used notations accepted in formula \eqref{rozv_fon-N}. From the structure of series \eqref{gth}
it is clear that in case of absence of correlations at the initial instant the correlations generated by
the dynamics of a system of hard spheres are completely determined by the cumulants of the groups of
operators \eqref{cumcp}.

\subsection{The Liouville hierarchy}
If $g_{s}^{0}\in L^{1}_{s},\,s\geq1,$ then for $t\in\mathbb{R}$ the sequence
of correlation functions \eqref{ghs} is a unique solution of the Cauchy problem of the weak formulation
of the Liouville hierarchy \cite{GerS},\cite{GP11}:
\begin{eqnarray}\label{vNh}
   &&\hskip-8mm\frac{\partial}{\partial t}g_{s}(t,x_1,\ldots,x_s)=\mathcal{L}^{\ast}_{s}g_{s}(t,x_1,\ldots,x_s)+\\
   &&\sum\limits_{\mathrm{P}:\,(x_1,\ldots,x_s)=X_{1}\bigcup X_2}\,\sum\limits_{i_{1}\in X_{1}}
      \sum\limits_{i_{2}\in X_{2}}\mathcal{L}_{\mathrm{int}}^{\ast}(i_{1},i_{2})
      g_{|X_{1}|}(t,X_{1})g_{|X_{2}|}(t,X_{2}), \nonumber\\
   \nonumber\\
 \label{vNhi}
   &&\hskip-8mm g_{s}(t,x_1,\ldots,x_s)\big|_{t=0}=g_{s}^{0}(x_1,\ldots,x_s),\quad s\geq1,
\end{eqnarray}
where ${\sum\limits}_{\mathrm{P}:\,(x_1,\ldots,x_s)=X_{1}\bigcup X_2}$ is the sum over all possible
partitions $\mathrm{P}$ of the set $(x_1,\ldots,x_s)$ into two nonempty mutually disjoint subsets
$X_1$ and $X_2$, and the operator $\mathcal{L}^{\ast}_{s}$ is defined on the subspace
$L^{1}_{0}\subset L^{1}_{\alpha}$ by formula \eqref{Lstar}. It should be noted that the Liouville
hierarchy \eqref{vNh} is the evolution recurrence equations set.

For $t\geq0$ we give a few examples of recurrence equations set \eqref{vNh} for a system of hard spheres:
\begin{eqnarray*}
   &&\hskip-8mm\frac{\partial}{\partial t}g_{1}(t,x_1)=-\langle p_1,\frac{\partial}{\partial q_1}\rangle g_{1}(t,x_1),\\
   &&\hskip-8mm\frac{\partial}{\partial t}g_{2}(t,x_1,x_2)=
     -\sum\limits_{j=1}^{2}\langle p_j,\frac{\partial}{\partial q_j}\rangle g_{2}(t,x_1,x_2)+ \sigma^2\int_{\mathbb{S}_{+}^2}d\eta\langle\eta,(p_{1}-p_{2})\rangle
        \big(g_2(t,q_1,p_{1}^\ast,\\
   &&q_{2},p_{2}^\ast)\delta(q_{1}-q_{2}+\sigma\eta)-g_2(t,x_1,x_2)\delta(q_{1}-q_{2}-\sigma\eta)\big)+\\
   &&\sigma^2\int_{\mathbb{S}_{+}^2}d\eta\langle\eta,(p_{1}-p_{2})\rangle
        \big(g_1(t,q_1,p_{1}^\ast)g_1(t,q_{2},p_{2}^\ast)\delta(q_{1}-q_{2}+\sigma\eta)-\\
   &&g_1(t,x_1)g_1(t,x_2)\delta(q_{1}-q_{2}-\sigma\eta)\big),
\end{eqnarray*}
where it was used notations accepted above in definition (\ref{Lint}).

We note that because the Liouville hierarchy \eqref{vNh} is the recurrence evolution equations
set, we can construct a solution of the Cauchy problem \eqref{vNh},\eqref{vNhi}, integrating
each equation of the hierarchy as the inhomogeneous Liouville equation. For example, as a result
of the integration of the first two equations of the Liouville hierarchy \eqref{vNh}, we obtain
the following equalities:
\begin{eqnarray*}
    &&\hskip-5mm g_{1}(t,x_1)=S^\ast_{1}(t,1)g_{1}^{0}(x_1),\\
    &&\hskip-5mm g_{2}(t,1,2)=S^\ast_{2}(t,1,2)g_{2}^0(x_1,x_2)+\\
    &&\int\limits_{0}^{t}dt_{1}S^\ast_{2}(t-t_{1},1,2)\mathcal{L}^\ast_{\mathrm{int}}(1,2)
       S^\ast_{1}(t_{1},1)S^\ast_{1}(t_{1},2)g_{1}^0(x_1)g_{1}^0(x_2).
\end{eqnarray*}
Then for the corresponding term on the right-hand side of the second equality, an analog of the
Duhamel equation holds
\begin{eqnarray*}
    &&\int\limits_{0}^{t}dt_{1}S^\ast_{2}(t-t_{1},1,2)\mathcal{L}^\ast_{\mathrm{int}}(1,2)
       S^\ast_{1}(t_{1},1)S^\ast_{1}(t_{1},2)=\\
    &&=-\int\limits_{0}^{t}dt_{1}\frac{d}{dt_{1}}\big(S^\ast_{2}(t-t_{1},1,2)
       S^\ast_{1}(t_{1},1)S^\ast_{1}(t_{1},2)\big)=\nonumber\\
    &&=S^\ast_{2}(t,1,2)-S^\ast_{1}(t,1)S^\ast_{1}(t,2)=\mathfrak{A}_{2}(t,1,2),\nonumber
\end{eqnarray*}
where $\mathfrak{A}_{2}(t)$ is the second-order cumulant of groups of operators \eqref{cumcp}.
As a result of similar transformations for $ s> 2 $, the solution of the Cauchy problem
\eqref{vNh},\eqref{vNhi}, constructed by an iterative procedure, is represented in the form
of expansions \eqref{ghs}.

We remark that a steady solution of the Liouville hierarchy \eqref{vNh} is a sequence of the
Ursell functions on the allowed configurations of a hard-sphere system, i.e.
$g^{(eq)}=(0,e^{-\beta \frac{p^2_1}{2}},0,\ldots,0,\ldots)$,
where
$\beta$ is a parameter inversely proportional to temperature.

We emphasize that the dynamics of correlations, that is, the fundamental equations \eqref{vNh}
describing the evolution of correlations of states, can be used as a foundation for describing
the evolution of states of a system of both a finite and an infinite number of hard spheres
instead of the Liouville equation for states \cite{GG21}-\cite{P62}.


\section{Processes of the propagation of correlations in a hard-sphere system}
Another approach to the description of states of hard-sphere systems of both finite and infinite
number of particles is can be formulated as in above by means of functions determined by the
cluster expansions of the reduced distribution functions. Such functions are interpreted as
reduced correlation functions of states (marginal or $s$-particle correlation functions
\cite{G17}-\cite{GP13}, or cumulants of marginals \cite{TSRS20},\cite{DS}).
On a microscopic scale, the macroscopic characteristics of fluctuations of observables are
directly determined by means of the reduced correlation functions.

The following also outlines the approach to the description of the evolution of states by means
of both reduced distribution functions and reduced correlation functions which is based on the
dynamics of correlations in a system of hard spheres governed by the Liouville hierarchy of
equations for a sequence of correlation functions.

\subsection{Reduced correlation functions}
Traditionally reduced correlation functions are introduced by means of the cluster expansions
of the reduced distribution functions similar to the cluster expansions of the probability
distribution functions \eqref{D_(g)N} and on the set of allowed configurations
$\mathbb{R}^{3n}\setminus \mathbb{W}_n$ they have the form:
\begin{eqnarray}\label{FG}
   &&\hskip-8mm F_{s}(t,x_1,\ldots,x_s)=
      \sum\limits_{\mbox{\scriptsize$\begin{array}{c}\mathrm{P}:(x_1,\ldots,x_s)=\bigcup_{i}X_{i}\end{array}$}}
      \prod_{X_i\subset \mathrm{P}}G_{|X_i|}(t,X_i), \quad s\geq1,
\end{eqnarray}
where ${\sum\limits}_{\mathrm{P}:(x_1,\ldots,x_s)=\bigcup_{i} X_{i}}$ is the sum over all possible
partitions $\mathrm{P}$ of the set $(x_1,\ldots,x_s)$ into $|\mathrm{P}|$ nonempty mutually disjoint
subsets $X_i\subset(x_1,\ldots,x_s)$. As a consequence of this, the solution of recurrence relations
\eqref{FG} represented through reduced distribution functions as follows:
\begin{eqnarray}\label{gBigfromDFB}
   &&\hskip-12mm G_{s}(t,x_1,\ldots,x_s)=
     \sum\limits_{\mbox{\scriptsize $\begin{array}{c}\mathrm{P}:(x_1,\ldots,x_s)=\bigcup_{i}X_{i}\end{array}$}}
     (-1)^{|\mathrm{P}|-1}(|\mathrm{P}|-1)!\,\prod_{X_i\subset \mathrm{P}}F_{|X_i|}(t,X_i), \\
   &&\hskip-12mm s\geq1,\nonumber
\end{eqnarray}
are interpreted as the functions that describe the correlations of states in a hard-sphere system.
The structure of expansions \eqref{gBigfromDFB} is such that the reduced correlation functions
can be treated as cumulants (semi-invariants) of the reduced distribution functions \eqref{RozvBBGKY}.

We note that the reduced correlation functions give an equivalent approach to the description
of the evolution of states of many hard spheres along with the reduced distribution functions.
Indeed, the macroscopic characteristics of fluctuations of observables are directly determined
by the reduced correlation functions on the microscopic scale \cite{BogLect},\cite{GP13}, for
example, the functional of the dispersion of an additive-type observable, i.e. the sequence $A^{(1)}=(0,a_{1}(x_1),\ldots,\sum_{i_{1}=1}^{n}a_1(x_{i_{1}}),\ldots)$, is represented by
the formula
\begin{eqnarray*}
    &&\hskip-8mm \langle(A^{(1)}-\langle A^{(1)}\rangle)^2\rangle(t)=
      \int_{\mathbb{R}^{3}\times\mathbb{R}^{3}}dx_{1}\,(a_1^2(x_1)-\langle A^{(1)}\rangle^2(t))G_{1}(t,x_1)+\\
    &&\hskip+12mm
      \int_{(\mathbb{R}^{3}\times\mathbb{R}^{3})^2}dx_{1}dx_{2}\,a_{1}(x_1)a_{1}(x_2)G_{2}(t,x_1,x_2),
\end{eqnarray*}
where $\langle A^{(1)}\rangle(t)=\int_{\mathbb{R}^{3}\times\mathbb{R}^{3}}dx_{1}\,a_{1}(x_1)G_{1}(t,x_1)$
is the mean value functional of an additive-type observable.

If $G(0)=(1,G_1^{0}(x_1),\ldots,G_s^{0}(x_1,\ldots,x_s),\ldots)$ is a sequence of reduced correlation
functions at initial instant, then the evolution of all possible states, i.e. the sequence
$G(t)=(1,G_{1}(t,x_1),\ldots,G_{s}(t,x_1,\ldots,x_s),\ldots)$ of the reduced correlation functions
$G_{s}(t),\,s\geq1$, is determined by the following series expansions \cite{G17}:
\begin{eqnarray}\label{sss}
    &&\hskip-12mm G_{s}(t,x_1,\ldots,x_s)=\\
    &&\hskip-8mm\sum\limits_{n=0}^{\infty}\frac{1}{n!}
        \,\int_{(\mathbb{R}^{3}\times\mathbb{R}^{3})^{n}}dx_{s+1}\ldots dx_{s+n}\,
        \,\mathfrak{A}_{1+n}(t;\{1,\ldots,s\},s+1,\ldots,s+n\mid G(0)), \nonumber\\
    &&\hskip-12mm s\geq1,\nonumber
\end{eqnarray}
where the generating operator $\mathfrak{A}_{1+n}(t;\{1,\ldots,s\},s+1,\ldots,s+n\mid G(0))$ of this
series is the $(1+n)th$-order cumulant of groups of nonlinear operators \eqref{ghs}:
\begin{eqnarray}\label{cc}
   &&\hskip-9mm\mathfrak{A}_{1+n}(t;\{1,\ldots,s\},s+1,\ldots,s+n\mid G(0))\doteq\\
   &&\sum\limits_{\mathrm{P}:\,(\{1,\ldots,s\},s+1,\ldots,s+n)=
      \bigcup_k X_k}(-1)^{|\mathrm{P}|-1}({|\mathrm{P}|-1})!
      \mathcal{G}(t;\theta(X_1)\mid\ldots \nonumber\\
    &&\mathcal{G}(t;\theta(X_{|\mathrm{P}|})\mid G(0))\ldots), \quad n\geq0,\nonumber
\end{eqnarray}
and where the composition of mappings \eqref{ghs} of the corresponding noninteracting groups of particles
was denoted by $\mathcal{G}(t;\theta(X_1)\mid \ldots\mathcal{G}(t;\theta(X_{|\mathrm{P}|})\mid G(0))\ldots)$,
for example,
\begin{eqnarray*}
    &&\hskip-5mm\mathcal{G}\big(t;1\mid\mathcal{G}(t;2\mid G(0))\big)=
        \mathfrak{A}_{1}(t,1)\mathfrak{A}_{1}(t,2)G^{0}_{2}(x_1,x_2),\\
    &&\hskip-5mm\mathcal{G}\big(t;1,2\mid\mathcal{G}(t;3\mid G(0))\big)=
        \mathfrak{A}_{1}(t,\{1,2\})\mathfrak{A}_{1}(t,3)G^{0}_{3}(x_1,x_2,x_3)+\\
    &&\hskip+5mm\mathfrak{A}_{2}(t,1,2)\mathfrak{A}_{1}(t,3)
       \big(G^{0}_{1}(x_1)G^{0}_{2}(x_2,x_3)+G^{0}_{1}(x_2)G^{0}_{2}(x_1,x_3)\big).
\end{eqnarray*}

We will adduce examples of expansions \eqref{cc}. The first order cumulant of the groups
of nonlinear operators \eqref{ghs} is the group of these nonlinear operators
\begin{eqnarray*}
     &&\mathfrak{A}_{1}(t;\{1,\ldots,s\}\mid G(0))=\mathcal{G}(t;1,\ldots,s\mid G(0)).
\end{eqnarray*}
In case of $s=2$ the second order cumulant of nonlinear operators \eqref{ghs}
has the structure
\begin{eqnarray*}
     &&\hskip-8mm \mathfrak{A}_{1+1}(t;\{1,2\},3\mid G(0))=\mathcal{G}(t;1,2,3\mid G(0))-
       \mathcal{G}\big(t;1,2\mid\mathcal{G}(t;3\mid G(0))\big)=\\
     && \mathfrak{A}_{1+1}(t,\{1,2\},3)G^{0}_{3}(1,2,3)+\\
     && \big(\mathfrak{A}_{1+1}(t,\{1,2\},3)-
        \mathfrak{A}_{2}(t,2,3)\mathfrak{A}_{1}(t,1)\big)G^{0}_{1}(x_1)G^{0}_{2}(x_2,x_3)+\\
     &&\big(\mathfrak{A}_{1+1}(t,\{1,2\},3)-
        \mathfrak{A}_{2}(t,1,3)\mathfrak{A}_{1}(t,2)\big)G^{0}_{1}(x_2)G^{0}_{2}(x_1,x_3)+\\
     && \mathfrak{A}_{1+1}(t,\{1,2\},3)G^{0}_{1}(x_3)G^{0}_{2}(x_1,x_2)+
        \mathfrak{A}_{3}(t,1,2,3)G^{0}_{1}(x_1)G^{0}_{1}(x_2)G^{0}_{1}(x_3),
\end{eqnarray*}
where the operator
\begin{equation*}
    \mathfrak{A}_{3}(t,1,2,3)=\mathfrak{A}_{1+1}(t,\{1,2\},3)-\mathfrak{A}_{2}(t,2,3)\mathfrak{A}_{1}(t,1)-
     \mathfrak{A}_{2}(t,1,3)\mathfrak{A}_{1}(t,2)
\end{equation*}
is the third-order cumulant \eqref{cumcp} of groups of operators \eqref{ad} of a system of hard spheres.

In the case of the initial state specified by the sequence of reduced correlation functions
$G^{(c)}=(0,G_1^{0},0,\ldots,0,\ldots)$, that is, in the absence of correlations between hard spheres
at the initial moment of time \cite{BPSh},\cite{Sh}, according to definition \eqref{cc}, on the allowed
configurations reduced correlation functions \eqref{sss} are represented by the following series expansions:
\begin{eqnarray}\label{mcc}
   &&\hskip-8mm G_{s}(t,x_1,\ldots,x_s)=\\
   &&\hskip-5mm \sum\limits_{n=0}^{\infty}\frac{1}{n!}
       \,\int_{(\mathbb{R}^{3}\times\mathbb{R}^{3})^{n}}dx_{s+1}\ldots dx_{s+n}\,\mathfrak{A}_{s+n}(t;1,\ldots,s+n)
       \prod_{i=1}^{s+n}G_1^{0}(x_i)\mathcal{X}_{\mathbb{R}^{3(s+n)}\setminus \mathbb{W}_{s+n}}, \nonumber\\
   &&\hskip-8mm  s\geq1, \nonumber
\end{eqnarray}
where the generating operator $\mathfrak{A}_{s+n}(t)$ is the $(s+n)th$-order cumulant \eqref{cumcp}
of groups of operators \eqref{ad}.

If $G(0)\in\oplus_{n=0}^{\infty}L^{1}_{n}$, then provided that
$\max_{n\geq1}\big\|G_n^{0}\big\|_{L^{1}_{n}}<(2e^{3})^{-1}$ \cite{G17}, for $t\in\mathbb{R}$ the sequence
of reduced correlation functions \eqref{sss} is a unique solution of the Cauchy problem of the hierarchy
of evolution nonlinear equations for hard spheres (for quantum systems known as the nonlinear BBGKY
hierarchy \cite{BogLect}):
\begin{eqnarray}\label{gBigfromDFBa}
   &&\hskip-8mm\frac{\partial}{\partial t}G_s(t,x_1,\ldots,x_s)=\mathcal{L}^{\ast}_{s}G_{s}(t,x_1,\ldots,x_s)+\\
   && \sum\limits_{\mathrm{P}:\,(x_1,\ldots,x_s)=X_{1}\bigcup X_2}\,\sum\limits_{i_{1}\in X_{1}}
      \sum\limits_{i_{2}\in X_{2}}\mathcal{L}_{\mathrm{int}}^{\ast}(i_{1},i_{2})
      G_{|X_{1}|}(t,X_{1})G_{|X_{2}|}(t,X_{2}))+\nonumber\\
   &&\int_{\mathbb{R}^{3}\times\mathbb{R}^{3}}dx_{s+1}
      \sum_{i=1}^{s}\mathcal{L}^{\ast}_{\mathrm{int}}(i,s+1)\big(G_{s+1}(t,x_1,\ldots,x_{s+1})+\nonumber\\
   && \sum_{\mbox{\scriptsize$\begin{array}{c}\mathrm{P}:(1,\ldots,s+1)=X_1\bigcup X_2,\\i\in
      X_1;s+1\in X_2\end{array}$}}G_{|X_1|}(t,X_1)G_{|X_2|}(t,X_2)\big),\nonumber\\ \nonumber\\
 \label{gBigfromDFBai}
   &&\hskip-8mmG_{s}(t,x_1,\ldots,x_s)\big|_{t=0}=G_{s}^{0}(x_1,\ldots,x_s), \quad s\geq1,
\end{eqnarray}
where the generators of these evolution equations are defined as in \eqref{Lstar},
and we used notations accepted in the Liouville hierarchy of equations \eqref{vNh}.

\subsection{On the description of states governed by the dynamics of correlations}
A definition equivalent to the definition \eqref{ms} of reduced distribution functions
can be formulated on the basis of the correlation functions \eqref{ghs} of systems of
a finite number of hard spheres, namely (see Appendix)
\begin{eqnarray}\label{FClusters}
    &&\hskip-12mm F_{s}(t,x_1,\ldots,x_s)\doteq\\
    &&\hskip-12mm \sum\limits_{n=0}^{\infty}\frac{1}{n!}\,
       \int_{(\mathbb{R}^{3}\times\mathbb{R}^{3})^{n}}dx_{s+1}\ldots dx_{s+n}\,
       g_{1+n}(t,\{x_1,\ldots,x_s\},x_{s+1},\ldots,x_{s+n}), \quad s\geq1,\nonumber
\end{eqnarray}
where the correlation functions of clusters of hard spheres $g_{1+n}(t), n\geq0,$ are
defined by the expansions:
\begin{eqnarray}\label{rozvNF-Nclusters}
    &&\hskip-8mm g_{1+n}(t,\{x_1,\ldots,x_s\},x_{s+1},\ldots,x_{s+n})=\\
    &&\hskip-5mm \sum\limits_{\mathrm{P}:\,(\{1,\ldots,s\},\,s+1,\ldots,s+n)=\bigcup_i X_i}
       \mathfrak{A}_{|\mathrm{P}|}\big(t,\{\theta(X_1)\},\ldots,\{\theta(X_{|\mathrm{P}|})\}\big)
       \prod_{X_i\subset \mathrm{P}}g_{|X_i|}^0(X_i),\nonumber\\
     &&\hskip-8mm n\geq0,\nonumber
\end{eqnarray}
and $\mathfrak{A}_{|\mathrm{P}|}(t)$ is the $|\mathrm{P}|th$-order cumulant \eqref{cumulantP}
of the groups of operators \eqref{rozv_fon-N}. The possibility of redefining of the reduced
distribution functions naturally arises as a result of dividing the series in expression \eqref{ms}
by the series of the normalization factor \cite{GP11}.

Since the correlation functions $g_{1+n}(t),\,n\geq0,$ are governed by the corresponding Liouville
hierarchy for clusters of hard spheres, the reduced distribution functions \eqref{FClusters} are
governed by the BBGKY hierarchy \eqref{BBGKY}.

We note that correlation functions of hard-sphere clusters expressed through correlation
functions of hard spheres \eqref{ghs} by the following relations:
\begin{eqnarray*}
  &&\hskip-8mm g_{1+n}(t,\{x_1,\ldots,x_s\},x_{s+1},\ldots,x_{s+n})=\\
  && \sum\limits_{\mathrm{P}:(\{x_1,\ldots,x_s\},\,x_{s+1},\ldots,x_{s+n})=\bigcup_i X_i}
      (-1)^{|\mathrm{P}|-1}(|\mathrm{P}|-1)!\times\\
  &&\prod_{X_i\subset \mathrm{P}}\,
      \sum\limits_{\mathrm{P'}:\,\theta(X_{i})=\bigcup_{j_i} Z_{j_i}}
      \prod_{Z_{j_i}\subset \mathrm{P'}}g_{|Z_{j_i}|}(t,Z_{j_i}), \quad n\geq0.
\end{eqnarray*}
In particular case $n=0$, i.e. the correlation function of a cluster of the $s$ hard spheres,
these relations take the form
\begin{eqnarray*}
  &&g_{1+0}(t,\{x_1,\ldots,x_s\})=\sum\limits_{\mathrm{P}:\,(x_1,\ldots,x_s)=\bigcup_{i} X_{i}}
      \prod_{X_{i}\subset \mathrm{P}}g_{|X_{i}|}(t,X_{i}).\nonumber
\end{eqnarray*}

Assuming as a basis an alternative approach to the description of the evolution of states of a
hard-sphere system within the framework of correlation functions \eqref{ghs}, then the reduced
correlation functions are defined by means of a solution of the Cauchy problem of the Liouville
hierarchy \eqref{vNh},\eqref{vNhi} as follows \cite{GP13},\cite{G17}:
\begin{eqnarray}\label{Gexpg}
   &&\hskip-8mm G_{s}(t,x_1,\ldots,x_s)\doteq\sum\limits_{n=0}^{\infty}\frac{1}{n!}\,
      \int_{(\mathbb{R}^{3}\times\mathbb{R}^{3})^{n}}dx_{s+1}\ldots dx_{s+n}
      \,\,g_{s+n}(t,x_1,\ldots,x_{s+n}),\\
   &&\hskip-8mm s\geq1,\nonumber
\end{eqnarray}
where the generating function $g_{s+n}(t,x_1,\ldots,x_{s+n})$ is defined by expansion
\eqref{gfromDFB}. Such a representation is derived as a result of the fact that the
reduced correlation functions are cumulants \eqref{gBigfromDFB} of reduced distribution
functions \eqref{FClusters}.

Since the correlation functions $g_{s+n}(t),\,n\geq0,$ are governed by the Liouville hierarchy
for hard spheres \eqref{vNh}, the reduced correlation functions defined as \eqref{Gexpg} are
governed by the nonlinear BBGKY hierarchy \eqref{gBigfromDFBa}.

We emphasize that $nth$ term of expansions \eqref{Gexpg} of the reduced correlation functions
are determined by the $(s+n)th$-particle correlation function \eqref{ghs} as contrasted to the
expansions of reduced distribution functions \eqref{FClusters} which are determined by the
$(1+n)th$-particle correlation function of clusters of hard spheres \eqref{rozvNF-Nclusters}.

In the absence of correlations of the states of hard spheres at the initial moment of time
on allowed configurations, the reduced correlation functions \eqref{sss} and the reduced
distribution functions are represented by expansions in the series \eqref{mcc} and \eqref{RozvBBGKY},
respectively. Consequently, the generator of these series expansions differs only in the order of
the cumulants of the groups of operators of hard spheres. As a result, the process of creating
correlations in a system of hard spheres is described by means of such reduced distribution
functions or reduced correlation functions.

Thus, as follows from the above, the cumulant structure of correlation function expansions
\eqref{rozvNF-Nclusters} or \eqref{ghs} induces the cumulant structure of series expansions
for reduced distribution functions \eqref{RozvBBGKY} and reduced correlation functions
\eqref{sss}, respectively, or other words, the evolution of the state of a system of an
infinite number of hard spheres is governed by the dynamics of correlations.


\section{On the description of correlations by means of the kinetic equations}

Further, an approach to the description of states by means of the state of a typical
particle of a system of many hard spheres is discussed, or in other words, foundations
are overviewed of describing the evolution of states by kinetic equations.

We shall consider systems which the initial state specified by a one-particle reduced
correlation (distribution) function, namely, the initial state specified by a sequence
of reduced correlation functions satisfying a chaos property stated above, i.e. by the
sequence $G^{(c)}=(0,G_1^{0},0,\ldots,0,\ldots)$. We remark that such an assumption about
initial states is intrinsic in kinetic theory of many-particle systems \cite{CGP97}-\cite{BPSh}.

Since the initial data $G^{(c)}$ is completely specified by the one-particle correlation
(distribution) function, the Cauchy problem of the nonlinear BBGKY hierarchy
\eqref{gBigfromDFBa},\eqref{gBigfromDFBai} is not completely well-defined  the Cauchy problem,
because the initial data is not independent for every unknown function of the hierarchy of
evolution equations. Therefore, the opportunity takes place to reformulate such a Cauchy
problem as a new Cauchy problem for the one-particle correlation function, with the independent
initial data and explicitly determined functionals of the solution of this Cauchy problem.
We formulate such a restated Cauchy problem and state functionals.

The following statement is true. In the case of the initial state specified by a one-particle
correlation function $G^{(c)}$ the evolution that described within the framework of the
sequence $G(t)=\left(1,G_1(t),\ldots,G_s(t),\ldots\right)$ of reduced correlation functions
\eqref{sss}, is also be described by the sequence
$G(t\mid G_{1}(t))=(1,G_1(t),G_2(t\mid G_{1}(t)),\ldots,G_s(t\mid G_{1}(t)),\ldots)$ of the
reduced (marginal) correlation functionals: $G_s(t,x_1,\ldots,x_s\mid G_{1}(t)),\,s\geq2$, with
respect to the one-particle correlation function $G_1(t)$ governed by the generalized Enskog
kinetic equation \cite{GG}.

A similar statement was proved in the article \cite{GG} for the states of a system of hard spheres
described in terms of the reduced distribution functions governed by the BBGKY hierarchy \eqref{BBGKY}.

In the case under consideration the reduced correlation functionals $G_s(t\mid G_{1}(t)),\,s\geq2$,
are represented with respect to the one-particle correlation function
\begin{eqnarray}\label{ske}
   &&\hskip-8mm G_{1}(t,x_1)=\\
   &&\hskip-5mm \sum\limits_{n=0}^{\infty}\frac{1}{n!}\,
      \int_{(\mathbb{R}^{3}\times\mathbb{R}^{3})^{n}}dx_{2}\ldots dx_{1+n}\,
      \mathfrak{A}_{1+n}(t, 1,\ldots, n+1)\prod_{i=1}^{n+1}G_{1}^{0}(x_i)
      \mathcal{X}_{\mathbb{R}^{3(n+1)}\setminus \mathbb{W}_{n+1}},\nonumber
\end{eqnarray}
where the generating operator $\mathfrak{A}_{1+n}(t)$ of this series is the $(1+n)th$-order
cumulant \eqref{cumcp} of the groups of operators \eqref{rozv_fon-N}, by the following series:
\begin{eqnarray}\label{f}
     &&\hskip-8mm G_{s}\bigl(t,x_1,\ldots,x_s\mid G_{1}(t)\bigr)=\\
     &&\sum_{n=0}^{\infty }\frac{1}{n!}\,
         \int_{(\mathbb{R}^{3}\times\mathbb{R}^{3})^{n}}dx_{s+1}\ldots dx_{s+n}\,
         \mathfrak{V}_{s+n}\bigl(t,1,\ldots,s+n\bigr)\prod_{i=1}^{s+n}G_{1}(t,x_i),\nonumber\\
     &&\hskip-8mm s\geq2.\nonumber
\end{eqnarray}
The generating operator $\mathfrak{V}_{s+n}(t),\,n\geq0$, of the $(s+n)th$-order of this series
is determined by the following expansion \cite{GG}:
\begin{eqnarray}\label{skrrc}
   &&\hskip-8mm\mathfrak{V}_{s+n}\bigl(t,1,\ldots,s,s+1,\ldots,s+n\bigr)=\\
   &&\hskip-8mm n!\,\sum_{k=0}^{n}\,(-1)^k\,\sum_{n_1=1}^{n}\ldots
       \sum_{n_k=1}^{n-n_1-\ldots-n_{k-1}}\frac{1}{(n-n_1-\ldots-n_k)!}\times\nonumber\\
   &&\hskip-8mm \hat{\mathfrak{A}}_{s+n-n_1-\ldots-n_k}(t,1,\ldots,s+n-n_1-\ldots-n_k)\times\nonumber\\
   &&\hskip-8mm \prod_{j=1}^k\,\sum\limits_{\mbox{\scriptsize$\begin{array}{c}
       \mathrm{D}_{j}:Z_j=\bigcup_{l_j}X_{l_j},\\
       |\mathrm{D}_{j}|\leq s+n-n_1-\dots-n_j\end{array}$}}\frac{1}{|\mathrm{D}_{j}|!}
       \sum_{i_1\neq\ldots\neq i_{|\mathrm{D}_{j}|}=1}^{s+n-n_1-\ldots-n_j}\,
       \prod_{X_{l_j}\subset \mathrm{D}_{j}}\,\frac{1}{|X_{l_j}|!}
       \hat{\mathfrak{A}}_{1+|X_{l_j}|}(t,i_{l_j},X_{l_j}),\nonumber
\end{eqnarray}
where $\sum_{\mathrm{D}_{j}:Z_j=\bigcup_{l_j} X_{l_j}}$ is the sum over all possible dissections
of the linearly ordered set $Z_j\equiv(s+n-n_1-\ldots-n_j+1,\ldots,s+n-n_1-\ldots-n_{j-1})$ on no
more than $s+n-n_1-\ldots-n_j$ linearly ordered subsets, the $(s+n)th$-order scattering cumulant
is defined by the formula
\begin{eqnarray*}
    &&\hskip-8mm\hat{\mathfrak{A}}_{s+n}(t,1,\ldots,s+n)\doteq
      \mathfrak{A}_{s+n}(t,1,\ldots,s+n)\mathcal{X}_{\mathbb{R}^{3(s+n)}\setminus \mathbb{W}_{s+n}}
      \prod_{i=1}^{s+n}\mathfrak{A}_{1}^{-1}(t,i),
\end{eqnarray*}
and notations accepted above were used. A method of the construction of reduced correlation
functionals \eqref{f} is based on the application of the so-called kinetic cluster expansions
\cite{GG} to the generating operators \eqref{cumcp} of series \eqref{mcc}. If
$\|G_{1}(t)\|_{L^{1}(\mathbb{R}^{3}\times\mathbb{R}^{3})}<e^{-(3s+2)}$, series (\ref{f})
converges in the norm of the space $L^{1}_{s}$ for arbitrary $t\in\mathbb{R}$ \cite{GG}.

We adduce simplest examples of generating operators \eqref{skrrc}:
\begin{eqnarray*}
   &&\hskip-8mm\mathfrak{V}_{s}(t,1,\ldots,s)=
      \mathfrak{A}_{s}(t,1,\ldots,s)\mathcal{X}_{\mathbb{R}^{3s}\setminus \mathbb{W}_{s}}
      \prod_{i=1}^{s}\mathfrak{A}_{1}^{-1}(t,i),\\
   &&\hskip-8mm\mathfrak{V}_{s+1}(t,1,\ldots,s,s+1)=\mathfrak{A}_{s+1}(t,1,\ldots,s+1)
      \mathcal{X}_{\mathbb{R}^{3(s+1)}\setminus \mathbb{W}_{s+1}}
      \prod_{i=1}^{s+1}\mathfrak{A}_{1}^{-1}(t,i)-\\
   &&\mathfrak{A}_{s}(t,1,\ldots,s)\mathcal{X}_{\mathbb{R}^{3s}\setminus \mathbb{W}_{s}}
   \prod_{i=1}^{s}\mathfrak{A}_{1}^{-1}(t,i)\times\\
   &&\sum_{j=1}^s\mathfrak{A}_{2}(t,j,s+1)\mathcal{X}_{\mathbb{R}^{6}\setminus \mathbb{W}_{2}}
   \mathfrak{A}_{1}^{-1}(t,j)\mathfrak{A}_{1}^{-1}(t,s+1).
\end{eqnarray*}

We note that reduced correlation functionals \eqref{f} describe all possible correlations
generated by the dynamics of many hard spheres in terms of a one-particle correlation function.

If $G_{1}^{0}\in L^{1}_1$, then for arbitrary $t\in\mathbb{R}$ one-particle correlation function
\eqref{ske} is a weak solution of the Cauchy problem of the generalized Enskog kinetic equation \cite{GG}
\begin{eqnarray}\label{gkec}
   &&\hskip-8mm\frac{\partial}{\partial t}G_{1}(t,x_1)=\mathcal{L}^{\ast}(1)G_{1}(t,x_1)+
      \int_{\mathbb{R}^{3}\times\mathbb{R}^{3}}dx_{2}\,
      \mathcal{L}_{\mathrm{int}}^{\ast}(1,2)G_{1}(t,x_1)G_{1}(t,x_2)+\\
   &&\hskip+8mm \int_{\mathbb{R}^{3}\times\mathbb{R}^{3}}dx_{2}\,
      \mathcal{L}_{\mathrm{int}}^{\ast}(1,2)G_{2}\bigl(t,x_1,x_2\mid G_{1}(t)\bigr),\nonumber\\
   \nonumber\\
 \label{gkeci}
   &&\hskip-8mm G_{1}(t,x_1)\big|_{t=0}=G_{1}^{0}(x_1),
\end{eqnarray}
where the first part of the collision integral in equation \eqref{gkec} has the Boltzmann--Enskog
structure, and the second part of the collision integral is determined in terms of the two-particle
correlation functional represented by series expansion \eqref{f} and it describes all possible
correlations which are created by hard-sphere dynamics and by the propagation of initial correlations
related to the forbidden configurations.

Indeed, by virtue of definitions \eqref{bLint},\eqref{Lstar} of the generator of the generalized Enskog
equation \eqref{gkec}, for $t>0$ the kinetic equation get the following explicit form
\begin{eqnarray*}
   &&\hskip-5mm\frac{\partial}{\partial t}G_{1}(t,x_1)=
      -\langle p_1,\frac{\partial}{\partial q_1}\rangle G_{1}(t,x_1)+\\
   &&\sigma^2\int_{\mathbb{R}^3\times\mathbb{S}_{+}^{2}}d p_{2}d\eta\,\langle\eta,(p_{1}-p_{2})\rangle
        \big(G_{1}(t,p_{1}^\ast,q_1)G_{1}(t,p_{2}^\ast,q_{1}-\sigma\eta,)-
\end{eqnarray*}
\begin{eqnarray*}
   &&G_{1}(t,x_1)G_{1}(t,p_2,q_1+\sigma\eta)\big)+\\
   &&\sigma^2\int_{\mathbb{R}^3\times\mathbb{S}_{+}^{2}}d p_{2}d\eta\,\langle\eta,(p_{1}-p_{2})\rangle
       \big(G_{2}\bigl(t,p_{1}^\ast,q_1,p_{2}^\ast,q_{1}-\sigma\eta\mid G_{1}(t)\bigr)-\\
   &&G_{2}\bigl(t,x_1,p_2,q_1+\sigma\eta\mid G_{1}(t)\bigr)\big).
\end{eqnarray*}

Thus, for the initial state specified by a one-particle correlation function, then all possible
states of a system of hard spheres can be described without any approximations within the framework
of a one-particle correlation function governed by non-Markovian kinetic equation \eqref{gkec}, and
of explicitly defined functionals \eqref{f} of its solution \eqref{ske}.


\section{On the low-density approximation of reduced correlation functions}

The conventional philosophy of the description of the kinetic evolution consists of the following.
If the initial state specified by a one-particle distribution function, then the evolution of
states can be effectively described by means of a one-particle distribution function governed
by the nonlinear kinetic equation in a suitable scaling limit \cite{G58},\cite{Bog}.

In the last decade, the Boltzmann--Grad limit (low-density scaling limit) \cite{G58},\cite{L75}
of the reduced distribution functions constructed by means of the theory of perturbations were
rigorously established in numerous papers, for example, in papers \cite{TSRS20},\cite{PS16},\cite{GG21}
and references therein.

Further, we consider a scheme for constructing the scaling asymptotic behavior of reduced correlation
functions \eqref{mcc} in the particular case of the Boltzmann--Grad limit in the case of the above-mentioned
initial state, which is specified by the scaled one-particle correlation function $G_{1}^{0,\epsilon},$
satisfying the condition:
\begin{eqnarray*}
    &&|G_{1}^{0,\epsilon}(x_1)|\leq ce^{\textstyle-\frac{\beta}{2}{p^{2}_1}},
\end{eqnarray*}
where $\epsilon>0$ is a scaling parameter (the ratio of the diameter $\sigma>0$ to the mean free path of
hard spheres), $\beta>0$ is a parameter and $c<\infty$ is some constant, and for $t\geq0$ the operator
$\mathcal{L}^{\ast}_{\mathrm{int}}$ in the dimensionless hierarchy of equations \eqref{gBigfromDFBa}
is scaled in such a way that
\begin{eqnarray*}
     &&\hskip-9mm \mathcal{L}_{\mathrm{int}}^\ast(j_{1},j_{2})f_{n}=
        \epsilon^2\int_{\mathbb{S}_{+}^2}d\eta\langle\eta,(p_{j_{1}}-p_{j_{2}})\rangle
        f_n(x_1,\ldots,p_{j_{1}}^*,q_{j_{1}},\ldots,\\
     &&p_{j_{2}}^*,q_{j_{2}},\ldots,x_n)\delta(q_{j_{1}}-q_{j_{2}}+\epsilon\eta)-
        f_n(x_1,\ldots,x_n)\delta(q_{j_{1}}-q_{j_{2}}-\epsilon\eta)\big),\nonumber
\end{eqnarray*}
where the notations similar to (\ref{Lint}) are used.

We emphasize that the states of a system of infinitely many hard spheres are described by sequences
of functions bounded with respect to the configuration variables \cite{CGP97} as it assumed above.

We will assume the existence of such Boltzmann--Grad limit of the reduced correlation function
$G_{1}^{0,\epsilon}$ in the sense of weak convergence
\begin{eqnarray}\label{asic1}
   &&\mathrm{w-}\lim_{\epsilon\rightarrow 0}\big(\epsilon^{2}G_{1}^{0,\epsilon}(x_1)-g_{1}^0(x_1)\big)=0.
\end{eqnarray}

Since the $nth$ term of series \eqref{mcc} for the $s$-particle correlation function is determined
by the $(s+n)th$-order cumulant of asymptotically perturbed groups of operators \eqref{ad}, then
on the finite time interval in the Boltzmann--Grad limit the property of the propagation of initial
chaos holds in the following sense:
\begin{eqnarray}\label{Gcid}
   &&\mathrm{w-}\lim\limits_{\epsilon\rightarrow 0}\epsilon^{2s}G_{s}(t,x_1,\ldots,x_s)=0,\quad s\geq2.
\end{eqnarray}

The equality \eqref{Gcid} is derived by the following assertions.

If $|f_{s}|\leq ce^{\textstyle-\frac{\beta}{2}{\sum^s_{i=1} p^{2}_i}}$, then for arbitrary finite time
interval for asymptotically perturbed first-order cumulant \eqref{cumcp} of the groups of operators \eqref{ad},
i.e. for the strongly continuous group \eqref{ad} the following equality takes place \cite{CGP97},\cite{PG90}
\begin{eqnarray*}
    &&\mathrm{w-}\lim\limits_{\epsilon\rightarrow 0}\big(S^{\ast}_{s}(t,1,\ldots,s)f_{s}-
        \prod\limits_{j=1}^{s}S^{\ast}_{1}(t,j)f_{s}\big)=0.
\end{eqnarray*}
Therefore, for the $(s+n)th$-order cumulant of asymptotically perturbed groups of operators \eqref{ad} the
following equalities true:
\begin{eqnarray*}\label{apg}
   &&\mathrm{w-}\lim\limits_{\epsilon\rightarrow 0}\frac{1}{\epsilon^{2n}}\,
     \mathfrak{A}_{s+n}(t,1,\ldots,s+n)f_{s+n}=0, \quad s\geq2.
\end{eqnarray*}

If equality (\ref{asic1}) holds for the initial one-particle correlation operator, then
in the case of $s=1$ for the series expansion (\ref{mcc}) the following equality is true
\begin{eqnarray*}
   &&\mathrm{w-}\lim\limits_{\epsilon\rightarrow 0}\big(\epsilon^{2}G_{1}(t,x_1)-g_{1}(t,x_1)\big)=0,
\end{eqnarray*}
where for arbitrary finite time interval the limit one-particle correlation function
$g_1(t,x_1)$ is represented by the series
\begin{eqnarray}\label{1mco}
   &&\hskip-8mm g_{1}(t,x_1)=\sum\limits_{n=0}^{\infty}\int\limits_0^tdt_{1}\ldots
      \int\limits_0^{t_{n-1}}dt_{n}\,\int_{(\mathbb{R}^{3}\times\mathbb{R}^{3})^{n}}dx_{2}\ldots dx_{1+n}\,
      S^{\ast}_{1}(t-t_{1},1)\times \\
   &&\mathcal{L}^{0,\ast}_{\mathrm{int}}(1,2)\prod\limits_{j_1=1}^{2}S^{\ast}_{1}(t_{1}-t_{2},j_1)\ldots
      \prod\limits_{i_{n}=1}^{n}S^{\ast}_{1}(t_{n}-t_{n},i_{n})\times \nonumber\\
   &&\sum\limits_{k_{n}=1}^{n}\mathcal{L}^{0,\ast}_{\mathrm{int}}(k_{n},n+1)
      \prod\limits_{j_n=1}^{n+1}S^{\ast}_{1}(t_{n},j_n)\prod\limits_{i=1}^{n+1}g_1^{0}(x_i).\nonumber
\end{eqnarray}
In this series expansion for $t\geq0$ the operator $\mathcal{L}_{\mathrm{int}}^{0,\ast}(j_{1},j_{2})$
is defined by the formula
\begin{eqnarray*}
     &&\hskip-9mm \mathcal{L}_{\mathrm{int}}^{0,\ast}(j_{1},j_{2})f_{n}
        \doteq\int_{\mathbb{S}_{+}^2}d\eta\langle\eta,(p_{j_{1}}-p_{j_{2}})\rangle
        \big(f_n(x_1,\ldots,p_{j_{1}}^*,q_{j_{1}},\ldots,p_{j_{2}}^*,q_{j_{2}},\ldots,x_n)-\\
     &&f_n(x_1,\ldots,x_n)\big)\delta(q_{j_{1}}-q_{j_{2}}),\nonumber
\end{eqnarray*}
where notations accepted in formula \eqref{bLint} are used.

Thus, we conclude that the limit one-particle correlation function (\ref{1mco}) is a weak solution of
the Cauchy problem of the Boltzmann kinetic equation:
\begin{eqnarray*}
  &&\frac{\partial}{\partial t}g_{1}(t,x_1)=\mathcal{L}^{\ast}(1)g_{1}(t,x_1)+
     \int_{\mathbb{R}^{3}\times\mathbb{R}^{3}}dx_{2}\,
     \mathcal{L}^{0,\ast}_{\mathrm{int}}(1,2)g_{1}(t,x_1)g_{1}(t,x_2),\\
     \nonumber\\
  &&g_{1}(t,x_1)|_{t=0}=g_{1}^0(x_1),
\end{eqnarray*}
or, if $t\geq0$, for a system of hard spheres the Boltzmann equation has the following explicit form
\begin{eqnarray*}\label{Boltz}
   &&\hskip-5mm\frac{\partial}{\partial t}g_{1}(t,x_1)=
        -\langle p_1,\frac{\partial}{\partial q_1}\rangle g_{1}(t,x_1)+\\
   &&\int_{\mathbb{R}^3\times\mathbb{S}_{+}^{2}}d p_{2}d\eta\,\langle\eta,(p_{1}-p_{2})\rangle
        \big(g_1(t,q_1,p_{1}^\ast)g_1(t,q_{1},p_{2}^\ast)-g_1(t,x_1)g_1(t,q_1,p_2)\big).\nonumber
\end{eqnarray*}

We remark that some other approaches to the derivation of kinetic equations, in particular, for a system
of many hard spheres with initial correlations, were developed in the works \cite{GG}-\cite{G-b20}. In
\cite{GG18}, a hierarchy of kinetic equations describing the evolution of the observables of a hard-sphere
system in the low-density limit is constructed.


\section{Conclusion}

This article dealt with a hard-sphere system of a non-fixed, i.e. arbitrary but finite
average number of identical hard spheres. The possible approaches to describing the
evolution of the states of a system of hard spheres using various modifications of probability
distribution functions were considered. One of these approaches allows one to describe the
evolution of both a finite and an infinite average number of hard spheres using reduced
distribution functions \eqref{RozvBBGKY} or reduced correlation functions \eqref{sss},
which are governed by the dynamics of correlations \eqref{ghs}.

Above it was established that the notion of cumulants \eqref{cumulantP} of the groups of operators
\eqref{rozv_fon-N} underlies non-perturbative expansions of solutions for the fundamental evolution
equations describing the evolution of the state of a hard-sphere system, namely of the Liouville
hierarchy \eqref{vNh} for correlation functions, of the BBGKY hierarchy \eqref{BBGKY} for reduced
distribution functions and of the nonlinear BBGKY hierarchy \eqref{gBigfromDFBa} for reduced
correlation functions, as well as it underlies the kinetic description of infinitely many hard
spheres \eqref{f}.

We emphasize that the structure of expansions for correlation functions \eqref{rozvNF-Nclusters},
in which the generating operators are the cumulants of the corresponding order \eqref{cumulantP}
of the groups of operators \eqref{rozv_fon-N} of hard spheres, induces the cumulant structure of
series expansions for reduced distribution functions \eqref{RozvBBGKY}, reduced correlation
functions \eqref{sss} and marginal correlation functionals \eqref{f}. Thus, in fact, the dynamics
of systems of infinitely many hard spheres is generated by the dynamics of correlations.

The origin of the microscopic description of the collective behavior of a hard-sphere system by
a one-particle correlation (distribution) function that is governed by the generalized Enskog
kinetic equation \eqref{gkec} was also considered. One of the advantages of such an approach
to the derivation of kinetic equations from underlying dynamics consists of an opportunity to
construct the kinetic equations with initial correlations, which makes it possible to describe
the propagation of initial correlations in the scaling limits \cite{GG21},\cite{GT}. Another
advantage of this approach is related to the problem of a rigorous derivation of the non-Markovian-type
kinetic equations on the basis of the hard-sphere dynamics, which make it possible to describe
the memory effects in many-particle systems with collisional dynamics.

In addition, it was established that in the particular case of a low-density approximation for
initial states specified by a one-particle correlation function the asymptotic behavior of the
constructed reduced correlation functions \eqref{mcc} is governed by the Boltzmann kinetic
equation with hard-sphere collisions.


\newpage

\addcontentsline{toc}{section}{Appendix}
\section*{Appendix}

The possibility of the description of the evolution of states based on the dynamics of correlations
\eqref{FClusters} or \eqref{Gexpg} occurs naturally in consequence of dividing the series in expression
\eqref{averageD} by the series of the normalizing factor, or other words, as a result of redefining of
mean value functional \eqref{averageD}.

To provide evidence of this statement, we will introduce the necessary concepts and
prove the validity of some equalities.
On sequences of functions $f,\widetilde{f}\in L^{1}_\alpha$ we define the $\ast$-product
    $$(f\ast\widetilde{f})_{|Y|}(Y)=\sum\limits_{Z\subset Y}\,f_{|Z|}(Z)
        \,\widetilde{f}_{|Y\backslash Z|}(Y\backslash Z),\eqno(\text{A}. 1)$$
where $\sum_{Z\subset Y}$ is the sum over all subsets $Z$ of the set $Y\equiv(x_1,\ldots,x_s)$.
Using the definition of the $\ast$-product (\text{A}.1)
, we introduce the mapping
${\mathbb E}\mathrm{xp}_{\ast}$ and the inverse mapping ${\mathbb L}\mathrm{n}_{\ast}$ on
sequences $h=(0,h_1(x_1),\ldots,h_n(x_1,\ldots,x_n),\ldots)$ of functions $h_n\in L^{1}_n$
by the expansions
   $$({\mathbb E}\mathrm{xp}_{\ast}\,h )_{|Y|}(Y)=\big(\mathbb{I}+
      \sum\limits_{n=1}^{\infty} \frac{h^{\ast n}}{n!}\big)_{|Y|}(Y)=\eqno(\text{A}. 2)$$
   $$\delta_{|Y|,0}+\sum\limits_{\mathrm{P}:\,Y=\bigcup_{i}X_{i}}\,
      \prod_{X_i\subset \mathrm{P}}h_{|X_i|}(X_i),$$
where we used the notations accepted in formula \eqref{D_(g)N}, $\delta_{|Y|,0}$ is the Kronecker
symbol, $\mathbb{I}=(1,0,\ldots,0,\ldots)$, and respectively,
   $$({\mathbb L}\mathrm{n}_{\ast}(\mathbb{I}+h))_{|Y|}(Y)=
       \big(\sum\limits_{n=1}^{\infty} (-1)^{n-1}\,\frac{h^{\ast n}}{n}\big)_{|Y|}(Y)=\eqno(\text{A}. 3)$$
   $$\sum\limits_{\mathrm{P}:\,Y=\bigcup_{i}X_{i}}(-1)^{|\mathrm{P}|-1}(|\mathrm{P}|-1)!\,
       \prod_{X_i\subset \mathrm{P}}h_{|X_i|}(X_i).$$
Therefore in terms of sequences of operators recursion relations \eqref{D_(g)N} are rewritten
in the form
\begin{eqnarray*}\label{DtoGcircledStar}
    &&D(t)={\mathbb E}\mathrm{xp}_{\ast}\,\,g(t),
\end{eqnarray*}
where $D(t)=\mathbb{I}+(0,D_1(t,x_1),\ldots,D_n(t,x_1,\ldots,x_n),\ldots)$. As a result, we get
\begin{eqnarray*}
    &&g(t)={\mathbb L}\mathrm{n}_{\ast}\,\,D(t).
\end{eqnarray*}

Thus, according to definition (\text{A}.1) of the $\ast$-product and mapping (\text{A}.3),
in the component-wise form solutions of recursion relations \eqref{D_(g)N} are represented by
expansions \eqref{gfromDFB}.

For arbitrary $f=(f_{0},f_{1},\ldots,f_{n},\ldots)\in L^{1}_\alpha$ and $Y\equiv(x_1,\ldots,x_s)$
we will define the linear mapping $\mathfrak{d}_{Y}:f\rightarrow \mathfrak{d}_{Y}f$, by the formula
    $$(\mathfrak{d}_{Y} f)_{n}(x_1,\ldots,x_n)\doteq f_{|Y|+n}(Y,x_{|Y|+1},\ldots,x_{|Y|+n}),
       \quad n\geq0.\eqno(\text{A}. 4)$$
For the set $\{Y\}$ consisting of the one element $Y=(x_1,\ldots,x_s)$, we have, respectively
    $$(\mathfrak{d}_{\{Y\}} f)_{n}(x_1,\ldots,x_n)\doteq
        f_{1+n}(\{Y\},x_{s+1},\ldots,x_{s+n}),\quad n\geq0.\eqno(\text{A}. 5)$$
On sequences $\mathfrak{d}_{Y}f$ and $\mathfrak{d}_{Y'}\widetilde{f}$ we introduce the $\ast$-product
\begin{eqnarray*}
    &&(\mathfrak{d}_{Y}f\ast\mathfrak{d}_{Y'}\widetilde{f})_{|X|}(X)\doteq
       \sum\limits_{Z\subset X}f_{|Z|+|Y|}(Y,Z)\,\widetilde{f}_{|X\backslash Z|+|Y'|}(Y',X\backslash Z),
\end{eqnarray*}
where $X,Y,Y'$ are the sets, which terms characterize clusters of hard spheres, and $\sum_{Z\subset X}$
is the sum over all subsets $Z$ of the set $X$. In particular case $Y=\emptyset,\,Y'=\emptyset$,
this definition reduces to definition (\text{A}.1).

For $f=(0,f_{1},\ldots,f_{n},\ldots),\,f_{n}\in L^{1}_n$, according to definitions of mappings
(\text{A}.2) and (\text{A}.5), the following equality holds
    $$\mathfrak{d}_{\{Y\}}\mathbb{E}\mathrm{xp}_{\ast}f=
      \mathbb{E}\mathrm{xp}_{\ast}f\ast\mathfrak{d}_{\{Y\}}f,\eqno(\text{A}.6)$$
and for mapping (\text{A}.4) respectively
\begin{eqnarray*}
    &&\mathfrak{d}_{Y}\mathbb{E}\mathrm{xp}_{\ast}f=
       \mathbb{E}\mathrm{xp}_{\ast} f\ast\sum\limits_{\mathrm{P}:\,Y=\bigcup_i X_{i}}
       \mathfrak{d}_{X_1}f\ast\ldots\ast \mathfrak{d}_{X_{|\mathrm{P}|}}f,
\end{eqnarray*}
where ${\sum\limits}_{\mathrm{P}:\,Y=\bigcup_i X_{i}}$ is the sum over all possible partitions
$\mathrm{P}$ of the set $Y\equiv(x_1,\ldots,x_s)$ into $|\mathrm{P}|$ nonempty mutually disjoint
subsets $X_i\subset Y$.

According to the definition
\begin{eqnarray*}
    &&(I,f)\doteq\sum\limits_{n=0}^{\infty}\frac{1}{n!}
        \int_{(\mathbb{R}^{3}\times\mathbb{R}^{3})^{n}}dx_{1}\ldots dx_{n}\,f_{n}(x_1,\ldots,x_n),
\end{eqnarray*}
where $I=(1,\ldots,1,\ldots)$, for sequences $f,\widetilde{f}\in L^{1}_\alpha$, the following
equality holds
    $$(I,f\ast \widetilde{f})=(I,f)(I,\widetilde{f}).\eqno(\text{A}.7)$$

In terms of mappings (\text{A}.4) and (\text{A}.5) the generalized cluster expansions of
solutions \eqref{rozv_fon-N} of a sequence of the Liouville equations
    $$D_{s+n}(t,Y,\,X\setminus Y)=\sum\limits_{\mathrm{P}:(\{Y\},\,X\setminus Y)=\bigcup_i X_i}\,
      \prod_{X_i\subset \mathrm{P}}g_{|X_i|}(t,X_i),\quad s\geq1, \eqno(\text{A}.8)$$
where $X\setminus Y\equiv(x_{s+1},\ldots,x_{s+n})$, take the form
    $$\mathfrak{d}_{Y}D(t)=\mathfrak{d}_{\{Y\}}{\mathbb E}\mathrm{xp}_{\ast}\,\,g(t).$$

Now let us prove the equivalence of the definition \eqref{ms} of the reduced distribution
functions and the definition of \eqref{FClusters} in the framework of the correlation dynamics.

In terms of mapping (\text{A}.4) the definition of reduced distribution functions
\eqref{ms} is written as follows:
$$F_{s}(t,Y)=(I,D(t))^{-1}(I,\mathfrak{d}_{Y}D(t)).$$
Using generalized cluster expansions (\text{A}.8), and as a consequence of equalities
(\text{A}.6),(\text{A}.7), we find
\begin{eqnarray*}
    &&(I,\mathfrak{d}_{Y}D(t))=
       (I,\mathfrak{d}_{\{Y\}}{\mathbb E}\mathrm{xp}_{\ast}\,\,g(t))=\\
    &&(I,\mathbb{E}\mathrm{xp}_{\ast}g(t)\ast\mathfrak{d}_{\{Y\}}g(t))
       =(I,\mathbb{E}\mathrm{xp}_{\ast}g(t))(I,\mathfrak{d}_{\{Y\}}g(t)).
\end{eqnarray*}
Taking into account that, according to the particular case $Y=\emptyset,$ of cluster
expansions (\text{A}.8), the equality holds
\begin{eqnarray*}
    &&(I,\mathbb{E}\mathrm{xp}_{\ast}g(t))=(I,D(t)),
\end{eqnarray*}
and as a result, we establish the following representation for the reduced distribution
functions:
$$F_{s}(t,Y)=(I,\mathfrak{d}_{\{Y\}}g(t)).$$
Therefore, in componentwise-form, we obtain relation \eqref{FClusters}.

We remind that the correlation functions of particle clusters in series \eqref{FClusters},
i.e. the functions $g_{1+n}(t,\{Y\},X\setminus Y),\,n\geq0$, are defined as solutions of
generalized cluster expansions (\text{A}.8), namely
\begin{eqnarray*}\label{gClusters}
    &&\hskip-8mm g_{1+n}(t,\{Y\},X\setminus Y)=
      \sum\limits_{\mathrm{P}:(\{Y\},\,X\setminus Y)=\bigcup_i X_i}
      (-1)^{|\mathrm{P}|-1}(|\mathrm{P}| -1)!\,\prod_{X_i\subset \mathrm{P}}D(t,X_i),\\
    &&\hskip-8mm s\geq1,\,n\geq0,\nonumber
\end{eqnarray*}
where the probability distribution function $D(t,X_i)$ is solution \eqref{rozv_fon-N} of
the Liouville equation \eqref{vonNeumannEqn}.

Thus, we have established relation \eqref{FClusters} between the reduced distribution functions and
correlation functions. In a similar way, the validity of relation \eqref{Gexpg} between the reduced
correlation functions defined by the cumulant expansions: $G(t)={\mathbb L}\mathrm{n}_{\ast}\,F(t)$,
and correlation functions, i.e. $G_{s}(t,Y)=(I,\mathfrak{d}_{Y}g(t))$, can be justified.

\end{document}